\documentclass[traditabstract]{aa} 
\usepackage{graphicx}
\usepackage{txfonts}
\usepackage{natbib}

\usepackage{color}
\usepackage[normalem]{ulem}

\bibpunct{(}{)}{;}{a}{}{,} 
\newcommand{\comet}{{\circ \! < \!\!\!\!\!\!\ - \!\!\!\! - }}

\begin{document}
   \title{
P/2010~A2 LINEAR\thanks{Based on observations collected 
at the Gemini North Observatory, Mauna Kea, Hawai`i, USA, program
 GN-2009B-DD-10 
at the European Southern Observatory, La Silla, Chile, program 184.C-1143(A), 
and at the University of Hawai`i 2.2-m telescope, Mauna Kea, Hawai`i, USA.
}
}
 \subtitle{I: An impact in the Asteroid Main Belt}

   \author{
     O.~R.~Hainaut \inst{1} \and 
     J.~Kleyna     \inst{2,8} \and 
     G.~Sarid      \inst{2,8} \and 
     B.~Hermalyn   \inst{3} \and 
     A.~Zenn       \inst{2} \and 
     K.~J.~Meech   \inst{2,8} \and 
     P.~{H.~Schultz}     \inst{3} \and 
     H.~Hsieh      \inst{2} \and 
     G.~Trancho    \inst{6,7} \and 
     J.~Pittichov\'a \inst{2} \and 
     B.~Yang       \inst{2,8}        
          }

   \institute{
     European Southern Observatory (ESO), Karl Schwarzschild Stra\ss
     e, 85\,748 Garching bei M\"unchen, Germany \\  \email{ohainaut@eso.org}
     \and
     Institute for Astronomy (IfA), University of Hawai`i, 2680 Woodlawn Drive, Honolulu, HI 96\,822, USA
     \and
     Geological Sciences-Brown University, 324 Brook Street,
     Providence RI 02\,912, USA     
     \and 
     Instituto de Astrof\'{\i}sica de Canarias, c/V\'{\i}a Lactea
      s/n, 38200 La Laguna, Tenerife, Spain, and Departamento de
      Astrof\'{\i}sica, Universidad de La Laguna, E-38205 La Laguna,
      Tenerife, Spain 
     \and 
      INAF - Osservatorio Astrofisico di Arcetri, Largo E. Fermi 5,
      Firenze 50125, Italy 
     \and 
     Gemini Observatory, Colina El Pino s/n, La Serena, Chile
     \and
     Giant Magellan Telescope, P.O. Box 90\,933, Pasadena, CA 91\,109, USA 
     \and     
     NASA Astrobiology Institute, USA
}

   \date{Received 25 September 2011; accepted 10 November 2011}

  \abstract{
\object{Comet P/2010~A2 LINEAR} is an object on an asteroidal orbit
within the inner Main Belt, therefore a good candidate for membership
with the Main Belt Comet family. It was observed with several
telescopes (ESO New Technology Telescope, La Silla, Chile; Gemini
North, Mauna Kea, Hawai`i; University of Hawai`i 2.2~m, Mauna Kea,
Hawai`i) from 14~Jan. until 19~Feb. 2010 in order to characterize and
monitor it and its very unusual dust tail, which appears almost fully
detached from the nucleus; the head of the tail includes two narrow
arcs forming a cross.
No evolution was observed during the span of the
observations. Observations obtained during the Earth orbital plane
crossing allowed an examination of the out-of-plane 3D structure of
the tail. The immediate surroundings of the nucleus were found
dust-free, which allowed an estimate of the nucleus radius of
80--90~m, assuming an albedo $p=0.11$ and a phase correction with
$G=0.15$ (values typical for S-type asteroids).
A model of the thermal evolution indicates that such a small nucleus
could not maintain any ice content for more than a few
million years on its current orbit, ruling out ice sublimation 
dust ejection mechanism. Rotational spin-up and
electrostatic dust levitations were also rejected, leaving an impact with a
smaller body as the favoured hypothesis. This is further supported by 
the analysis of the tail structure.
Finston-Probstein dynamical dust modelling indicates the tail was
produced by a single burst of dust emission. More advanced models
(described in detail in a companion paper),
independently indicate that this burst populated a hollow cone with a
half-opening angle $\alpha\sim 40\degr$ and with an ejection velocity
$v_{\rm max}\sim 0.2$~m~s$^{-1}$, where the small dust grains fill the
observed tail, while the arcs are foreshortened sections of the burst
cone. 
The dust grains in the tail are measured to have radii between
$a=1$--20~mm, with a differential size distribution proportional to
$a^{-3.44\pm0.08}$. 
The dust contained in the tail is estimated to at least $8\times
10^8$~kg, which would form a sphere of 40~m radius (with a density
$\rho=3\,000$~kg~m$^{-3}$ and an albedo $p=0.11$ typical of S-type
asteroids).
Analysing these results in the framework of crater physics, we
conclude that a gravity-controlled crater would have grown up to $\sim
100$~m radius, i.e. comparable to the size of the body. The
non-disruption of the body suggest this was an oblique impact.
}

   \maketitle
%
\section{Introduction}

Habitability within our solar system is determined by the distribution
of water and volatiles, yet the origin of this distribution is
currently a fundamental unresolved planetary science issue
\citep{KC03_habitable}. There are three leading scenarios for the
origin of terrestrial planetary water, including: ($i$) nebular gas
adsorption on micron-sized dust grains inside the snow line
\citep[][the distance from the Sun outside of which water the
  temperature is low enough for water ice to condense]{Mur+08_water},
($ii$) chemical reactions between an early hydrogen envelope and oxides in a
magma ocean \cite{GI07_DHratio, GI08_ocean}, or ($iii$) delivery by
volatile-rich planetesimals (asteroids or comets) formed beyond the
snow line \citep{Mor+00_water}. The first two processes probably
contributed to, but may be unable to account for all of Earth's water
\citep{Mot+07_water}.  
Within the broader context of icy bodies, main belt comets (MBCs), a
newly discovered class of ``comets having stable orbits completely
confined to the main asteroid belt'' \citep{HsiehJ06_MBC}, present a
subclass of particular significance to the history of terrestrial
water and other important volatiles.

In addition to possibly P/2010~A2 (LINEAR), there are currently five
known MBCs: 133P/Elst-Pizarro, 176P/LINEAR, 238P/Read, C/2008 R1
(Garradd) and P/2010 R2 (La Sagra).  The main belt comets are defined
\citep{HsiehJ06_MBC} by ($i$) a semi-major axis that is less than
Jupiter's, ($ii$) Tisserand parameters significantly greater than 3
---meaning they are dynamically decoupled from Jupiter, like ordinary
asteroids \citep{Vaghi73, Kresak80}--- and ($iii$) mass loss with a
cometary appearance. As comets in near-circular orbits within the
asteroid belt, these objects likely still harbor nebular water frozen
out from beyond the primordial ``snow line'' \citep{Encrenaz08,
  GaraudL07_snow} of the young solar system.  Dynamical simulations
suggest they probably formed in-situ \citep{Haghighipour09_MBC} at a
different temperature from either comets or the asteroidal reservoir
that has been sampled through our meteorite collections.  Thus,
understanding their chemistry can provide unique insight into the
distribution of volatiles in the early stages of planet formation.

P/2010 A2 (LINEAR) was discovered by the LINEAR project on 7~Jan.,
2010 \citep{CBET2114} and was described as ``a headless comet with a
straight tail, and no central condensation'' \citep{IAUC9105}. On
11--12 Jan., observers at the WYNN 3.5-m and at the 2.5-m Nordic
Optical Telescope reported that the object had a asteroidal-like body
$\sim$ 150--200m in diameter which was connected to the tail by an
unresolved light bridge \citep{IAUC9109}. Based on the orbital
semi-major axis and on the Tisserand parameter, $T_J=3.6$, Jewitt
\citep{IAUC9109} concluded that P/2010~A2 was a new MBC. With the
smallest perihelion distance of any of the MBCs ($q=2.29$~AU), these
elements suggested a membership in the Flora family. Flora is a large
asteroid family of $\sim$500 members which can be broken up into many
sub-families or clans and which are broadly compositionally consistent
with space-weathered S-type asteroidal spectra \citep{Flo+98}.  Jewitt
\citep{IAUC9109} and Licandro \citep{CBET2134} suggested that the
location of the nucleus outside the coma might be the consequence of
an impact.

Is P/2010~A2 a genuine, sublimating comet (whose activity was possibly
triggered by an impact), or is the tail the signature of an impact (or
another alternative process), with a dispersion of the dust ejecta but
no sublimation? If it can be demonstrated that the object is a comet,
it would indicate that at least one object in the inner asteroid belt
still contains volatile material. The water snow line in the
proto-solar nebula is estimated to have been around 2.5~AU from the
Sun \citep{Jon+90} or possibly even as close as 1.5~AU
\citep{Lec+06,Mac+10,Min+11}. Initially parent bodies beyond the snow
line are believed to be mixtures of water ice and silicates, which are
then heated due to the radioactive decay of $^{26}$Al within 1--2
million years after nebular collapse \citep{Krot+06}.
\citet{Grimm+89} find that once the water is liquid, it is consumed by
hydration reactions, preferentially in the interior
\citep{Cohen+00,Wilson+99}, possibly leaving ice in the outer layers.
In the inner asteroid belt, it is likely that most of that water has
been converted in hydrated material observed on S-type asteroids
\citep{Riv+02}. MBCs may be the frozen components of the outer edges
of asteroid parent bodies that have survived the age of the solar
system.

Four of the MBCs are located in the outer main belt,
where asteroids with hydrated material are less common, suggesting
water ice could still exist.  The orbital elements of the MBCs are
listed in Table~\ref{tab:ele} and displayed in
Fig.~\ref{fig:ele}. 133P/Elst-Pizarro, the first MBC discovered, is a
member of the Themis family; 176P/LINEAR was found in a survey
targeting objects with orbits similar to that of 133P/Elst-Pizarro
\citep{HHH09}.  238P/Read, however, was discovered serendipitously
(i.e. not within a MBC-dedicated program) on
a similar orbit. A fourth MBC, P/2008~R1 Garradd, was discovered
serendipitously in the central region of the main belt. The fifth MBC,
P/2010~R2 La Sagra, was also discovered serendipitously 
 with a
semi-major axis similar to that of 133P/Elst-Pizarro
\citep{Mar+10}. More recently, asteroid (596)~Scheila presented a
comet-like appearance \citep{Lar10}. \citet{Bod+11} and \citet{Jew+11}
conclude however that this dust cloud was likely caused by the impact
of a small asteroid on Scheila. The survival of an icy body in the
inner asteroid belt would therefore give exciting constraints on the
evolution of water in the belt. It would also raise fundamental
questions on how to shield water ice in that area of the solar system
in a rather small body.
\begin{figure}
\caption{(not available on astroPh) The orbital elements of P/2010~A2 (green square to the left),
  of the 5 known MBCs (circles, 133P and 238P are
  indistinguishable) and of the numbered main belt asteroids (small
  dots). When available, the proper elements were used (source: the
  Asteroid Dynamics Site, http://hamilton.dm.unipi.it/astdys/). The
  inclination of P/2010~R2 La Sagra is off scale; its $a$ is marked by
  a triangle on the top of the plot; $i=21.39$ .
}
\label{fig:ele}
\end{figure}

\begin{figure}[h] 
{\bf a.} \includegraphics[width=8.8cm]{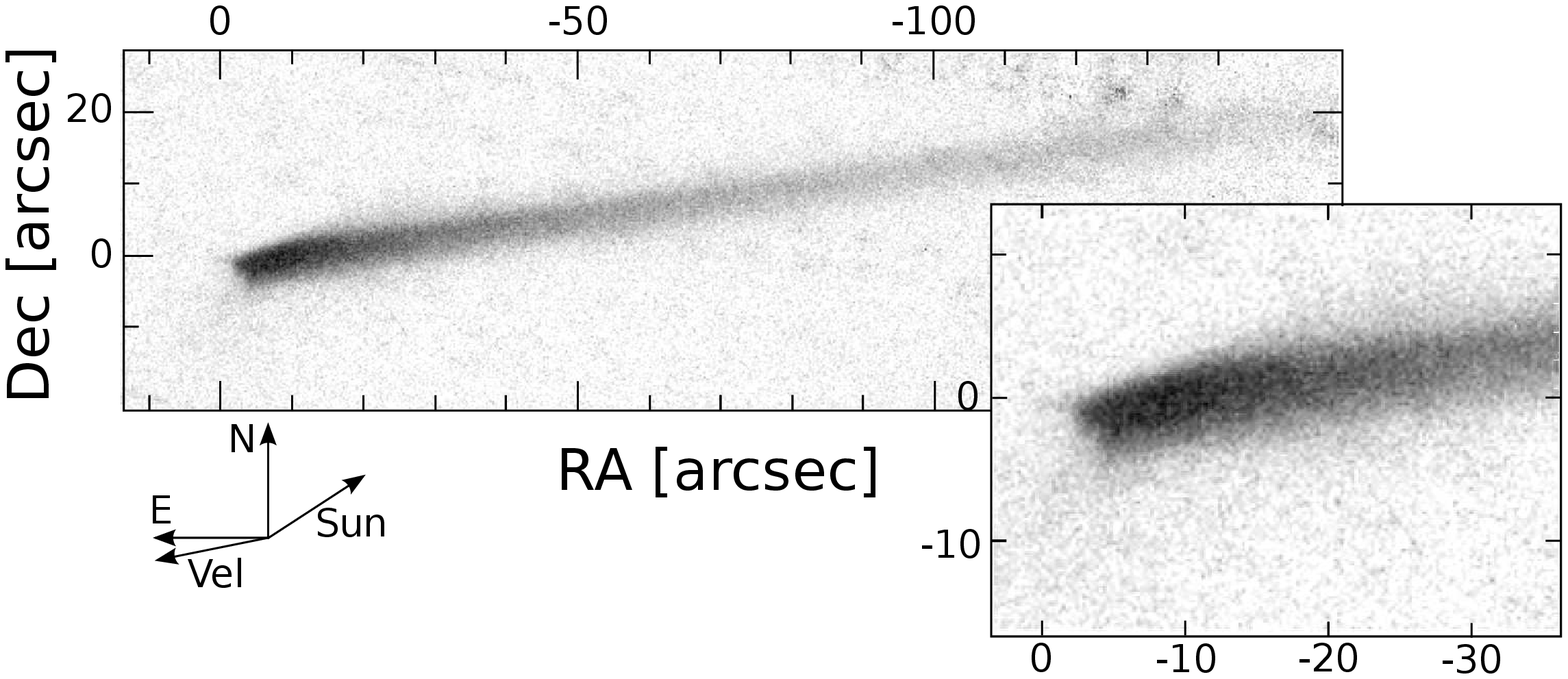}\\
{\bf b.} \includegraphics[width=8.8cm]{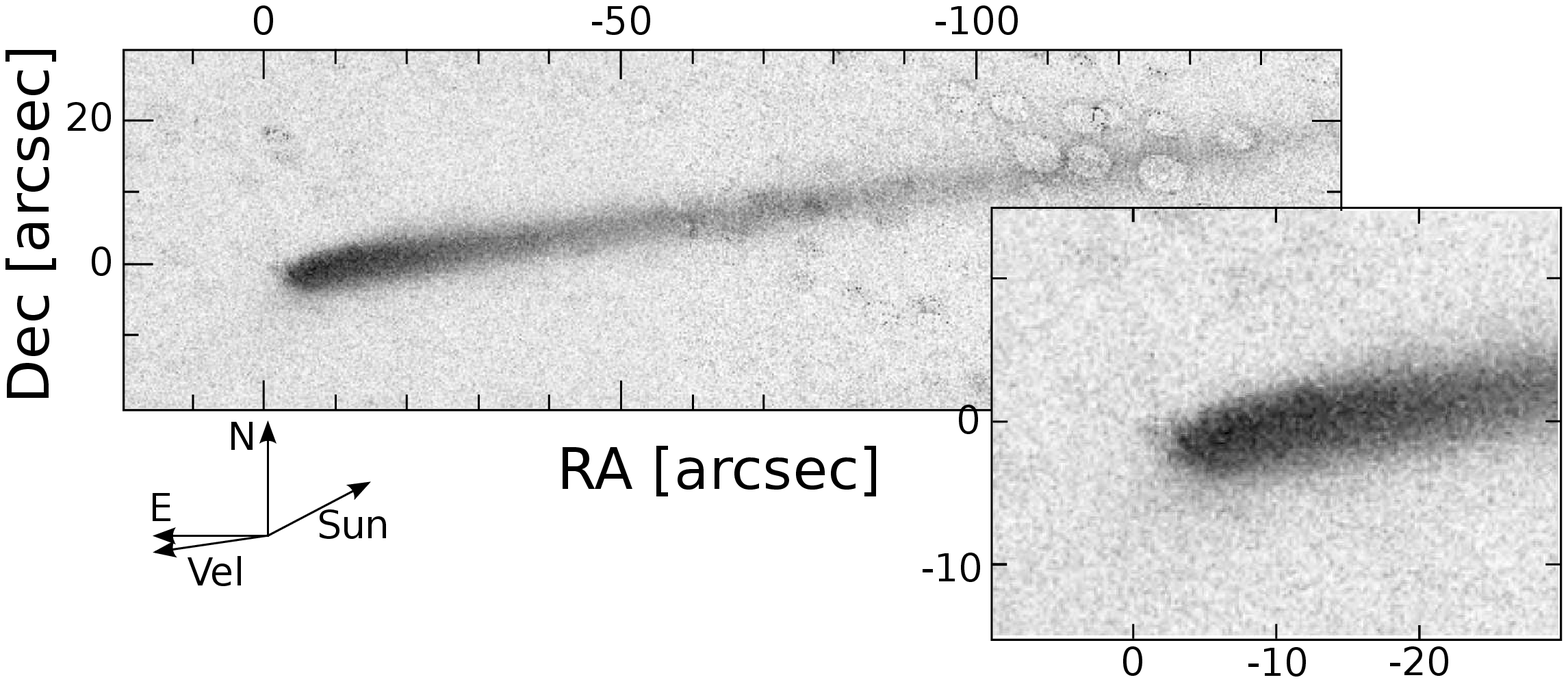}\\
{\bf c.} \includegraphics[width=8.8cm]{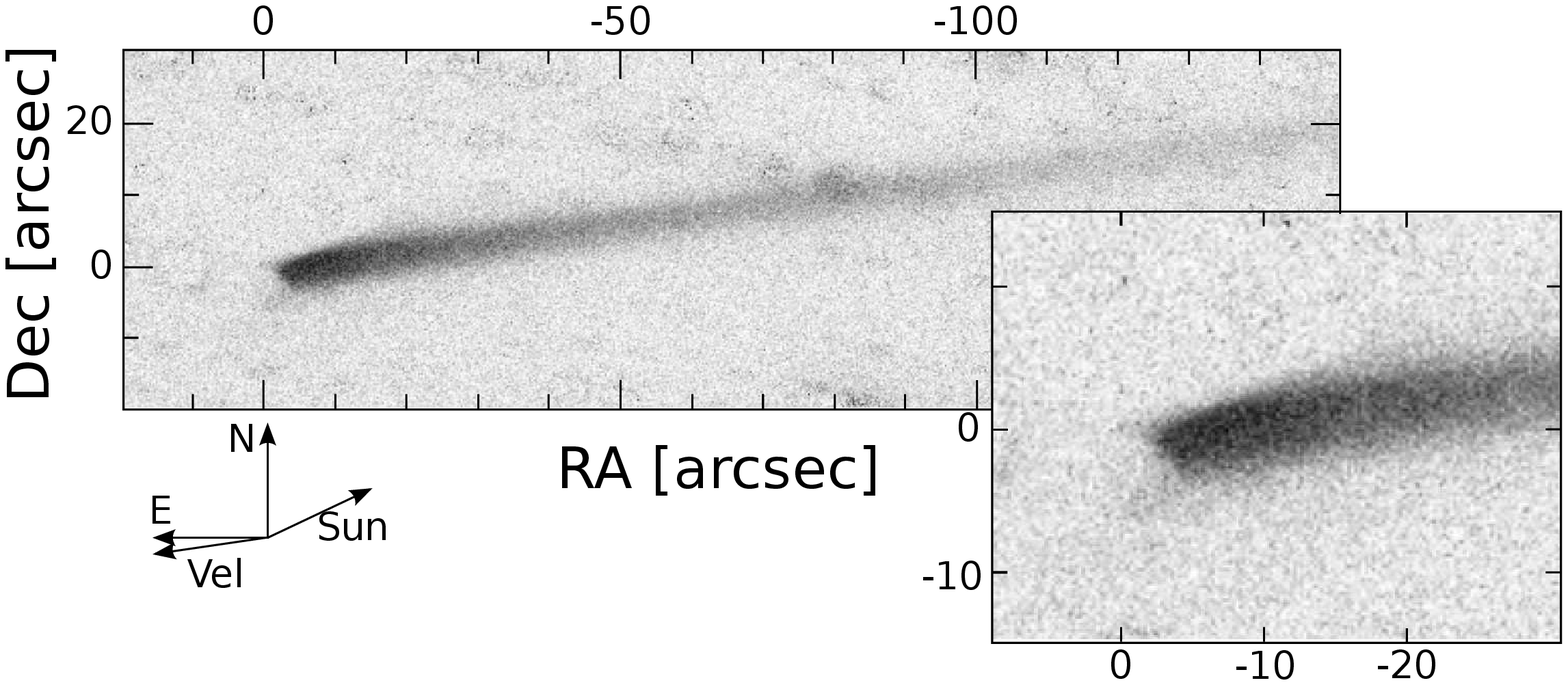}\\
{\bf d.} \includegraphics[width=8.8cm]{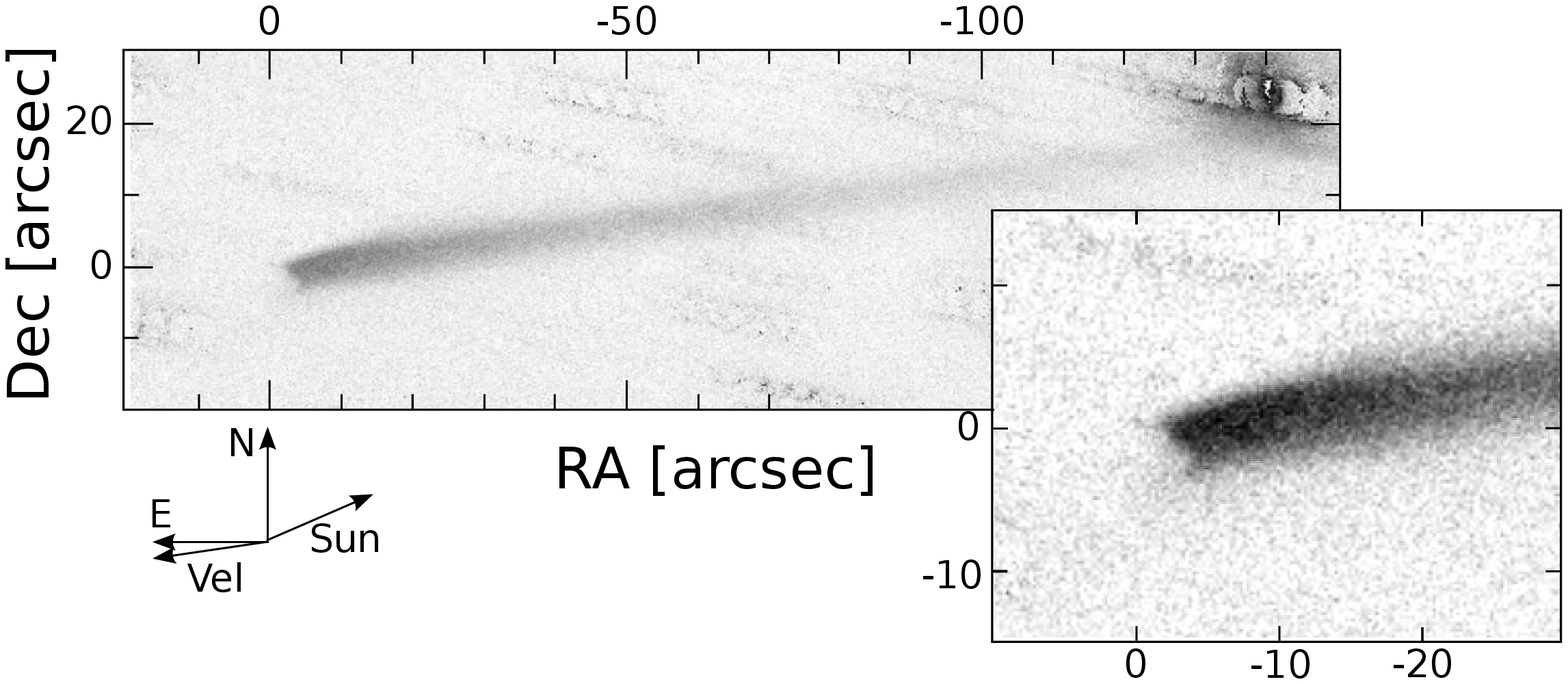}\\
{\bf e.} \includegraphics[width=8.8cm]{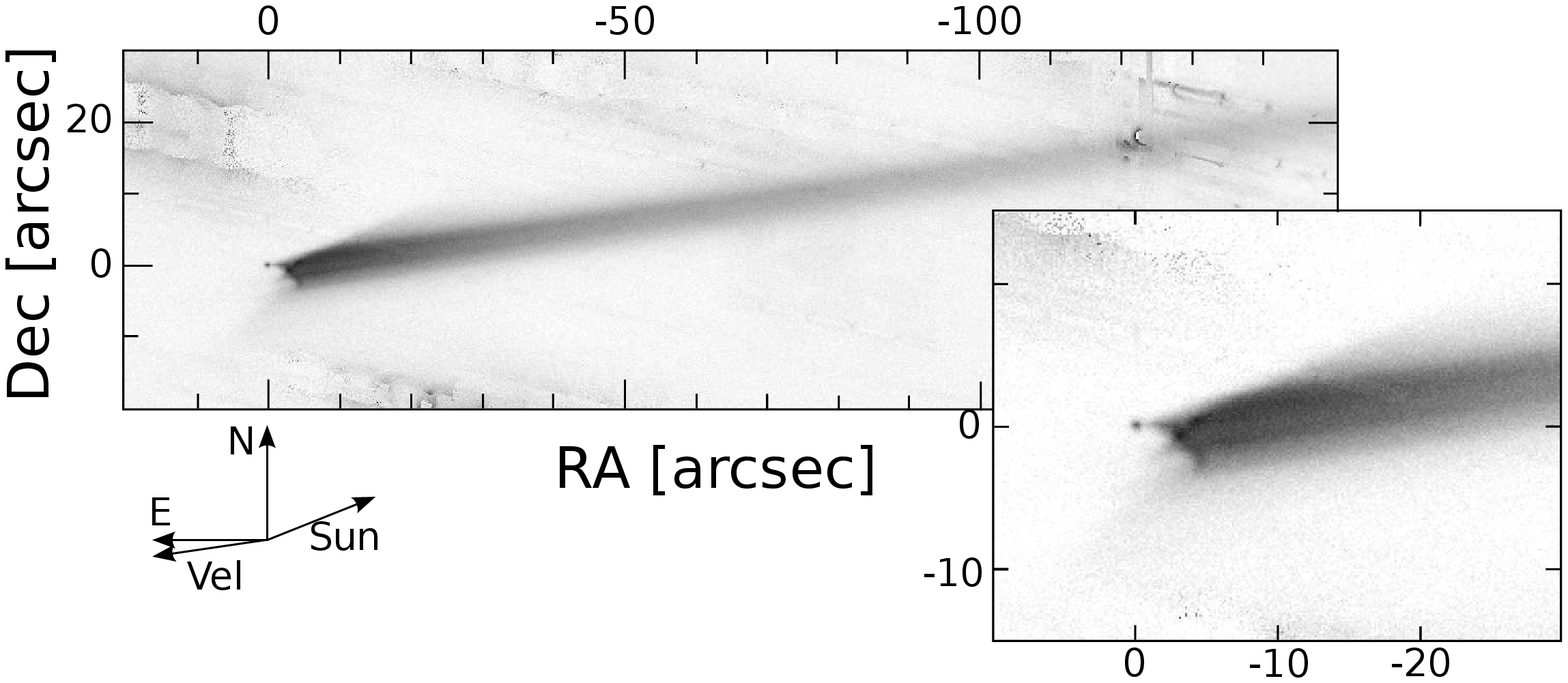}
\caption{P/2010~A2, images from 
  UT~14.2 (a), 
    16.2 (b), 
    17.2 (c), 
    18.2 (d) Jan. 2010 using the NTT, 
    UT~19.5 Jan. 2010 using GN (e).  
  The linear gray scale covers the range of (0--3) $\times 10^{-8}
  Af$, a proxy for the amount of dust present (see
  Section~\ref{sec:dust}). Each panel includes a 40$\times 30''$ zoomed
  portion to show details of the inner structure. The positions of N
  and E are indicated, as are the anti-solar direction and the
  heliocentric velocity vector.  } 
     \label{fig:image}%
\end{figure}
\addtocounter{figure}{-1}
\begin{figure}[h] 
{\bf f.}  \includegraphics[width=8.8cm]{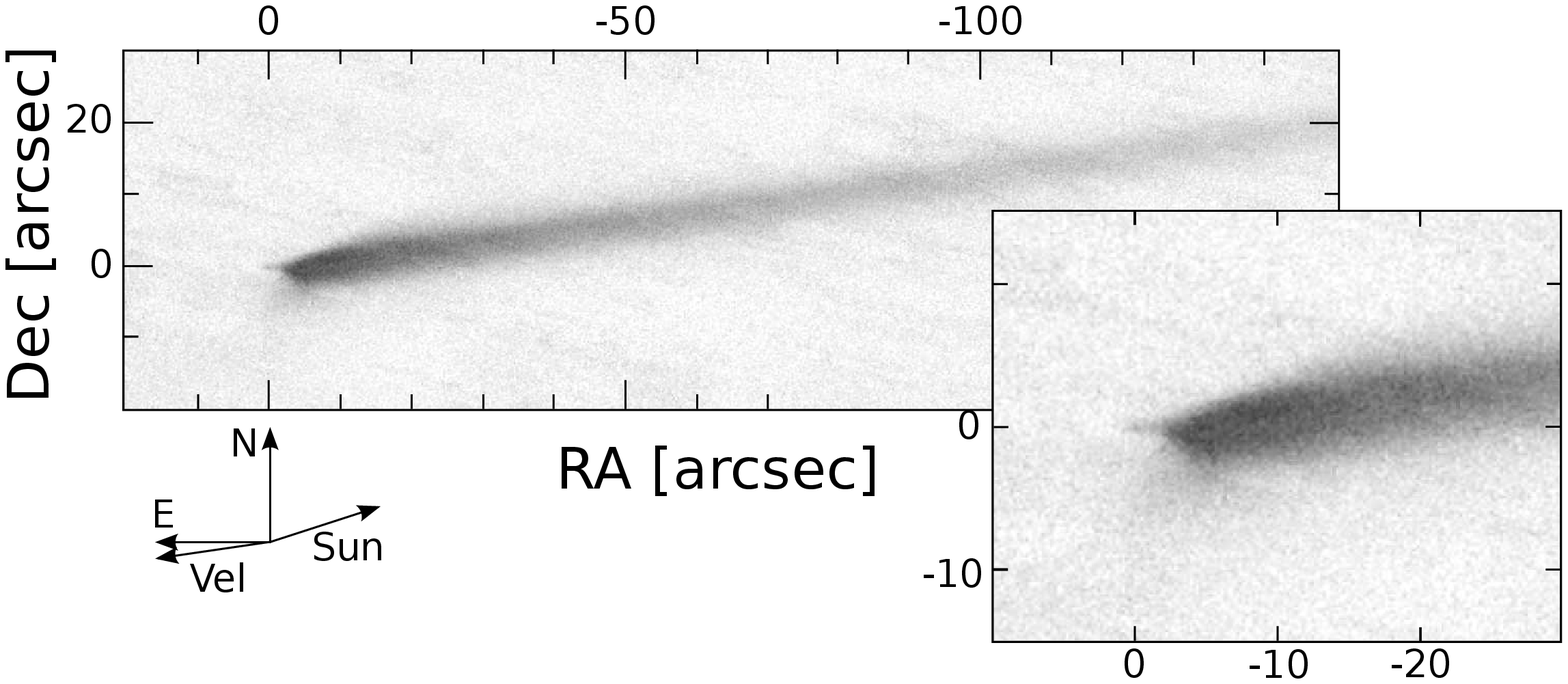}\\
{\bf g.}  \includegraphics[width=8.8cm]{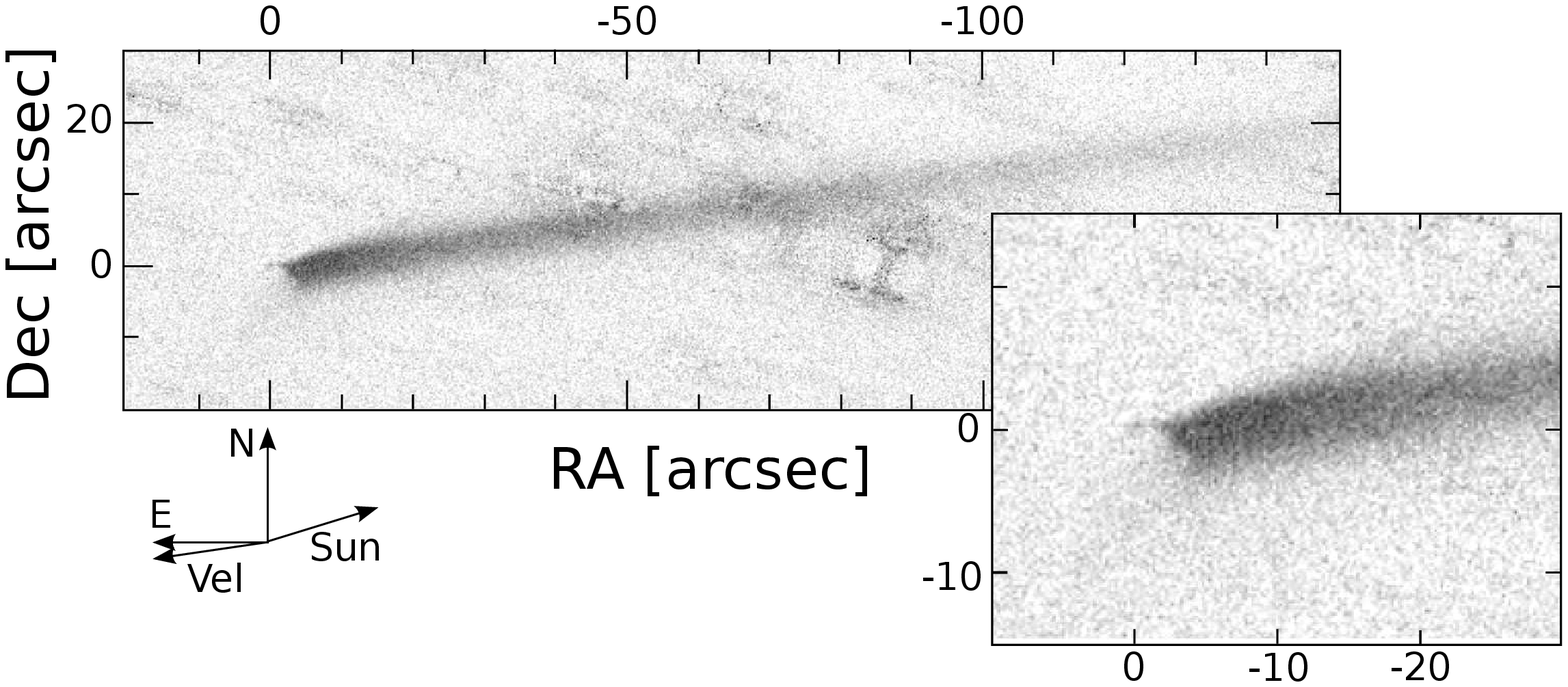}\\
{\bf h.}  \includegraphics[width=8.8cm]{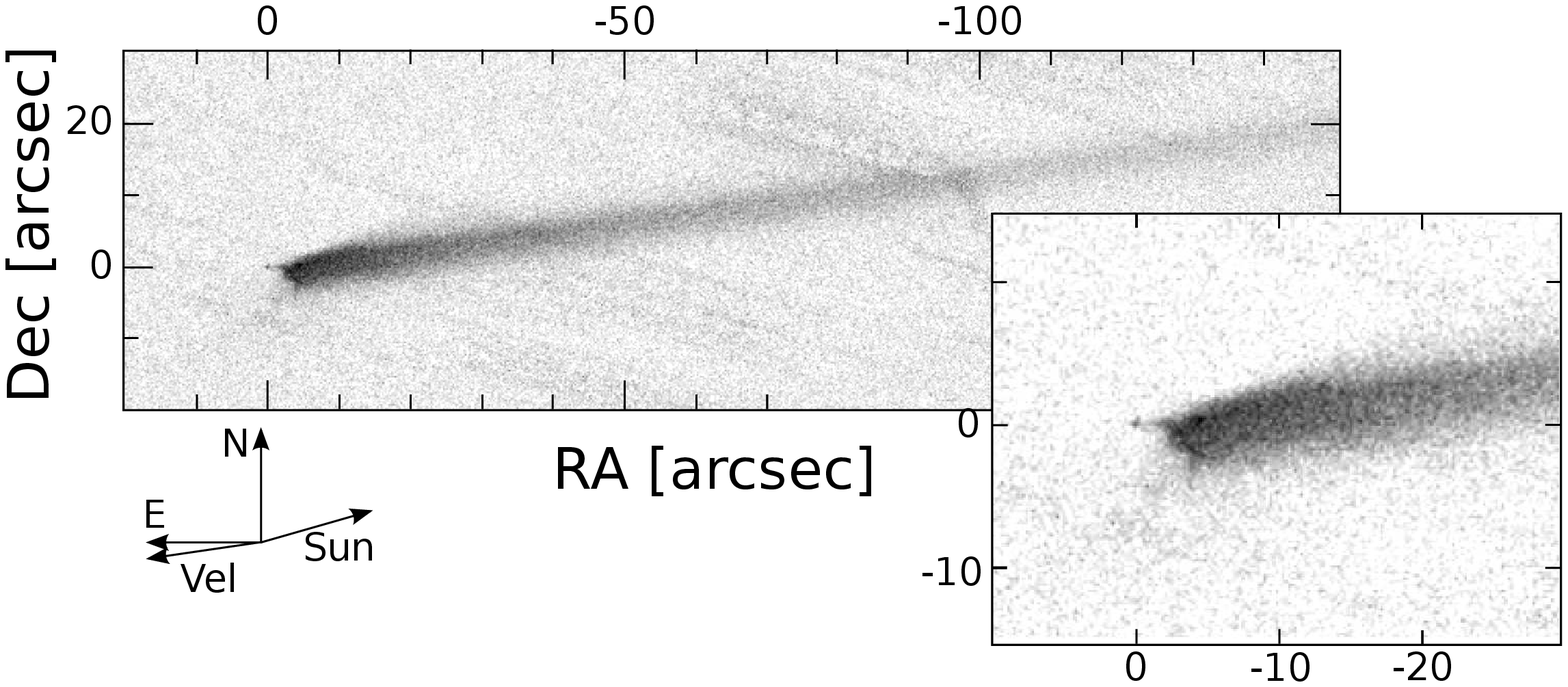}\\
{\bf i.}  \includegraphics[width=8.8cm]{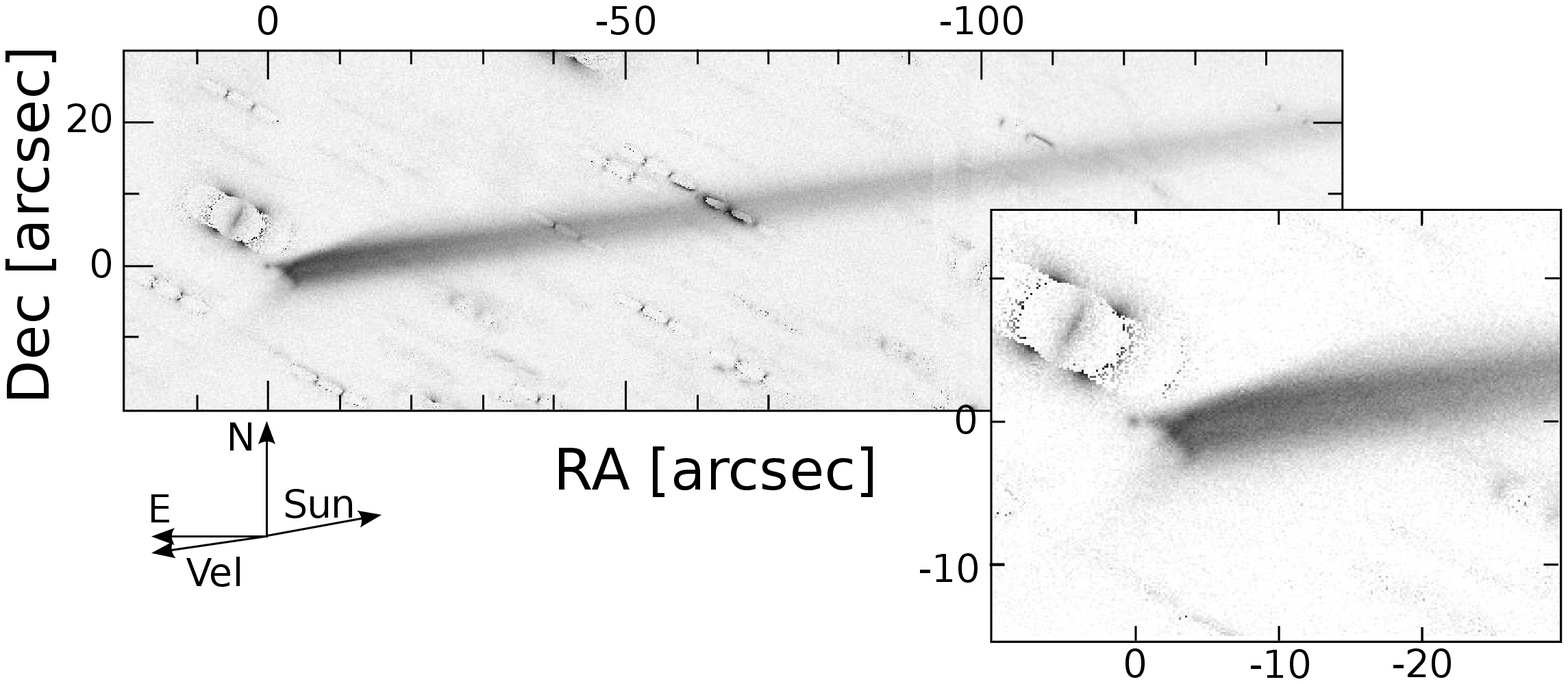}\\
{\bf j.}  \includegraphics[width=8.8cm]{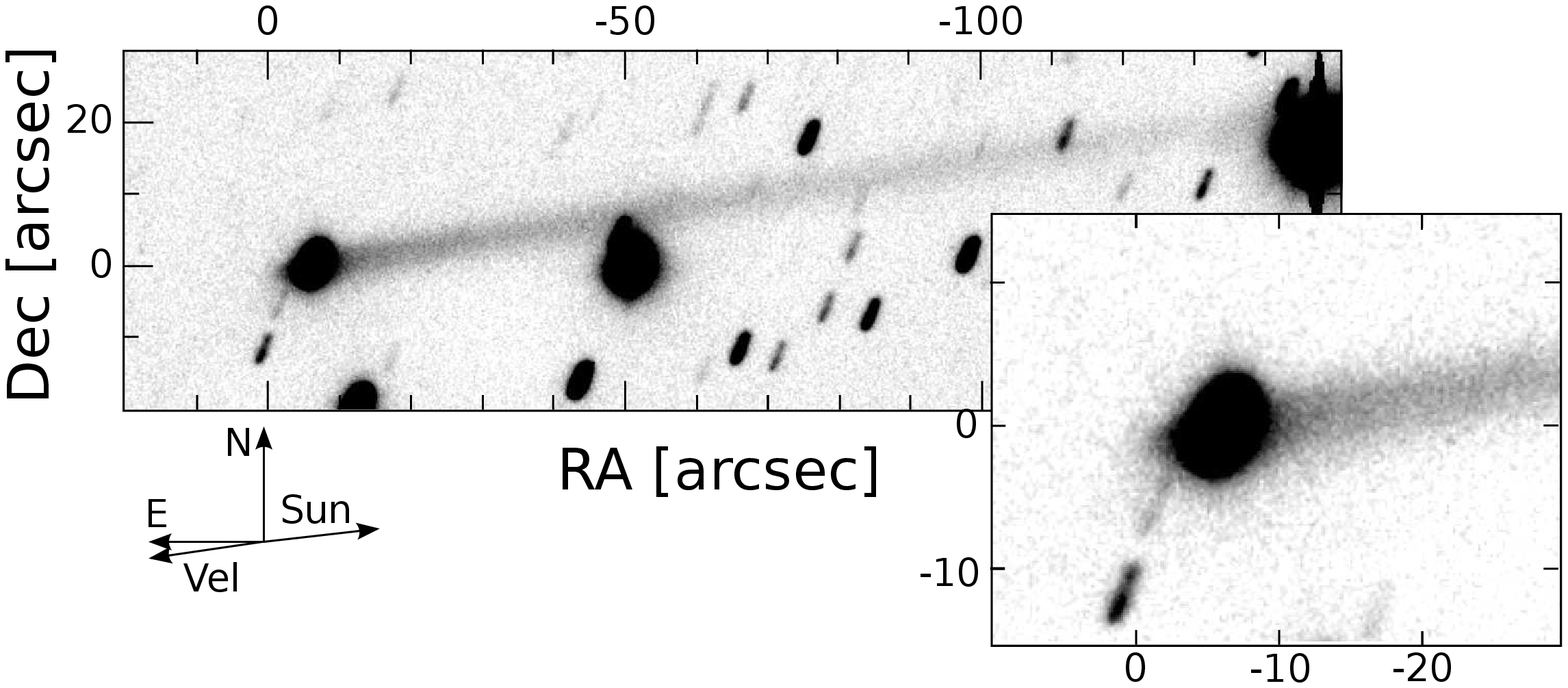}
  \caption{continued. P/2010~A2, images from 
UT~22.3 (f), 23.4 (g), 25.4 (h) Jan. 2010 using UH~2.2-m,
UT~2.3 Feb. 2010 using GN (i),
UT~19.5 Feb. 2010 using UH~2.2-m (j).} 
\end{figure}

\citet{Mor+10} presented observations obtained with the GTC, WHT and
NOT on La Palma, which they modeled with an extended period of
water-driven cometary activity. \citet{Jew+10} acquired a series of
HST images; from the orientation and geometry of the tail, they favor
the disruption of an asteroid (either by collision or spin up) in
Feb.--Mar. 2009. Finally, \citet{Sno+10} secured observation from the
Rosetta spacecraft. Thanks to the position of the space probe, they
could observe the object with a very different geometry (but with a
much more modest resolution), from which they concluded that the
observed dust tail was caused by an impact.

In order to investigate the process that generated the observed dust
tail, we acquired deep images of P/2010~A2 over various epochs. The
observations are presented in Section~\ref{sec:obs}. The analysis and
modelling of the nucleus and dust are described in
Section~\ref{sec:ana}.  In Section~\ref{sec:con}, we discuss the
conclusions and summarize the results.  A companion paper
\citep{PAPER2} is devoted to the details of the dust models developed
for this study, and a second follow-up article \citep{PAPER3} focuses
on the characteristics of the impact process and of the nucleus based on
the interpretation of the dust as an impact plume.

\section{Observations}\label{sec:obs}

The telescopes and instruments used are described below. The epoch,
geometric circumstances and a log of the observations are listed in
Table~\ref{tab:obs}.

In all cases, the observations were acquired as series of fairly short
individual exposures obtained while tracking at the non-sidereal
motion rates of the comet, and offsetting the telescope position
between exposures.

\subsection{Telescopes and Instruments}

\subsubsection{New Technology Telescope}
The observations were performed on the ESO 3.56m New Technology
Telescope (NTT) on La Silla, with the ESO Faint Object Spectrograph
and Camera (v.2) instrument (EFOSC2)
\citep{1984Msngr..38....9B,2008Msngr.132...18S}, through Bessel $B$,
$V$, $R$, and Gunn $i$ filters, using the ESO\#40 detector, a
2k$\times$2k thinned, UV-flooded Loral/Lesser CCD, which was read in a
2$\times$2 bin mode resulting in $0\farcs 24$ pixels, and a $4\farcm 1$ field
of view.


\subsubsection{Gemini North Telescope}

The observations were acquired in queue-mode on the 
8.1-m Gemini North (GN) Telescope on Mauna-Kea, using Gemini Multi-Object
Spectrograph (GMOS) \citep{2004PASP..116..425H}, through a SDSS $r'$
filter \citep{1996AJ....111.1748F}. The detector mosaic is composed of three
2k$\times$4k EEV CCDs arranged in a row, resulting in an un-vignetted
field of view of $5'$. The detectors were binned $2\times2$
resulting in a $0\farcs 145$ projected pixel size.

\subsubsection{Univertity of Hawai`i}
The observations were performed on the University of Hawai`i (UH)
2.2-m telescope on Mauna-Kea, using the Tektronix 2k$\times$2k CCD
camera \citep{UHCCD}. The projected pixel size is $0\farcs 22$, and
the field of view $7\farcm 5$. Kron-Cousin $B$, $V$, $R$, $I$ and SDSS
$r'$ filters were used.

\begin{table}
\caption{Elements of P/2010~A2 and the five known MBCs}     
\label{tab:ele}      
\begin{tabular}{lrrrr}
\hline
Object                & $a$ [AU] & $e$   & $i$ & $T_J$ \\
\hline
\hline
176P/(118401) LINEAR        &  3.217 & 0.150 &  1.352 & 3.172 \\
133P/(7968) Elst-Pizarro    &  3.164 & 0.154 &  1.370 & 3.185 \\
238P/Read (P/2005~U1)          &  3.165 & 0.253 &  1.266 & 3.153 \\
P/2008~R1 Garradd           &  2.726 & 0.342 & 15.903 & 3.216 \\
P/2010~R2 La Sagra          &  3.099 & 0.154 & 21.394 & 3.203 \\
\hline
P/2010A2 LINEAR             &  2.291 & 0.124 &  5.255 & 3.588 \\
\hline
\end{tabular}
\end{table}

\begin{table*}
\caption{Observation Log}     
\label{tab:obs}      
\begin{tabular}{lrrrrrrp{1.2cm}crlcp{2.3cm}}
\hline
 Obs. date    & $r^1$ &$\Delta^2$ &$\alpha^3$&PsAng$^4$& PsAMV$^5$ & PlAng$^6$& Tel-Ins$^7$  & Who$^8$& \multicolumn{2}{c}{Obs.$^9$}&Seeing&  Comments \\
 UT 2010        & [AU] & [AU]& [deg]& [deg] & [deg]& [deg]    &      &     &   &   &[arcsec]     & \\
\hline\hline                           
 14.2 Jan. & 2.013 & 1.040 &  5.6 & 123.3 & 280.1 & -2.2 & NTT& OH    & 1200&B, &  1.6 & Poor seeing\\
&&&&&&& EFOSC2&& 600& V,\\ 
&&&&&&&&& 2400& R,\\
&&&&&&&&& 800& i\\ 
 16.2 Jan. & 2.014 & 1.045 &  6.7 & 117.7 & 279.8 & -2.0 & NTT& HH    &  900& B,  &  1.4 & Poor seeing\\
&&&&&&& EFOSC2&& 600& V\\
&&&&&&&&&  1500& R,\\
&&&&&&&&&  600 & i \\
 17.2 Jan. & 2.014 & 1.048 &  7.2 & 115.5 & 279.7 & -2.0 & NTT& HH    &  600& B,  &  1.1 &    \\
&&&&&&& EFOSC2&& 600& V\\
&&&&&&&&& 1200 &R,\\
&&&&&&&&& 600 & i  \\
 18.2 Jan.& 2.015 & 1.051 &  7.8 & 113.6 & 279.6 & -1.9 & NTT EFOSC2& HH   &                3600& R         & 0.9--1.6 & \\
 19.5 Jan. & 2.015 & 1.055 &  8.5 & 111.5 & 279.4 & -1.8 & GN GMOS   & OH/GT &                3000& r$'$        & 0.5--0.7 & Excellent seeing.\\
 22.3 Jan. & 2.016 & 1.066 & 10.0 & 107.9 & 279.1 & -1.5 & UH Tek    & BY    &               1800 &B & 0.5-0.8 & \\
&&&&&&& Tek&& 1750& V\\
&&&&&&&&& 7250& R,\\
&&&&&&&&& 1800& I \\ 
 23.4 Jan.& 2.017 & 1.071 & 10.6 & 106.8 & 279.0 & -1.4 & UH Tek    & BY    &                3000 &R         & 1.0  & \\
 25.3 Jan. & 2.018 & 1.080 & 11.7 & 105.2 & 278.8 & -1.3 & UH Tek    & BY    &                4000& r$'$        & 0.7  & \\
 02.3 Feb.& 2.022 & 1.125 & 15.6 & 100.4 & 278.3 & -0.6 & GN GMOS   & OH/GT &                3000& r$'$        & 0.5   & \\
 19.5 Feb. & 2.032 & 1.262 & 22.3 &  96.3 & 278.3 &  0.7 & UH Tek    & JP    &                1500 &R         & 1.2 & Star on object\\
\hline                  
\end{tabular}
{\bf Notes:}
1, 2: Helio- and geocentric distances;
3: Solar phase angle;
4: Position angle of the extended Sun-Target radius vector; 
5: Position angle of the negative of the target's heliocentric
velocity vector;
6: Angle between observer and target orbital plane, measured from target;
7: Telescope and instrument: NTT= New Technology Telescope, GN= Genini North, UH= UH 2.2-m;
8: Observer initials: OH= O.~Hainaut, HH= H.~Hsieh, GT= G.~Trancho, BY= B.~Yang, JP= J.~Pittichov\'a
9: Total on-target exposure time [sec], and filter.

\end{table*}

\subsection{Data Processing}
\label{sec:dataproc}
The images were corrected for instrumental signature by subtracting a
master bias frame and dividing by a master flat-field frame obtained
from twilight sky images. Because of the large angular size of the
object, we have not applied second order, ``super-flatfield''
technique. In the case of the Gemini images, the basic processing was
performed using the GMOS tasks from the Gemini IRAF Package (version
1.9) , which also corrects for the geometric distortion and merges the
three chips into single images. The frames were flux calibrated using
nightly images of \citet{landolt92} fields obtained over a range of
airmasses.

In order to create deep composite images to assess the extent of the
dust, offsets between images were measured and corrected for using
series of field stars.  The motion of the comet was then compensated
for using either the linear ephemeris rates, or a full astrometric
solution to compute the offset between expected positions of the comet
from JPL's Horizon ephemerides. The frames were shifted and then combined
using either a median combination or an average rejecting the pixels
deviating the median value, in order to remove the stars and background
objects as well as detector defects and cosmic ray hits. In some
cases, individual stars were masked and rejected during the
combination. The resulting $R$-band (or equivalent) images for each
run are presented in Fig.~\ref{fig:image}. A zoomed portion of the
head of the object is included in each panel in order to show the
details of the inner structure.

\section{Analysis} \label{sec:ana}
\subsection{Description}\label{sec:description}

\begin{figure} 
\includegraphics[width=8.8cm]{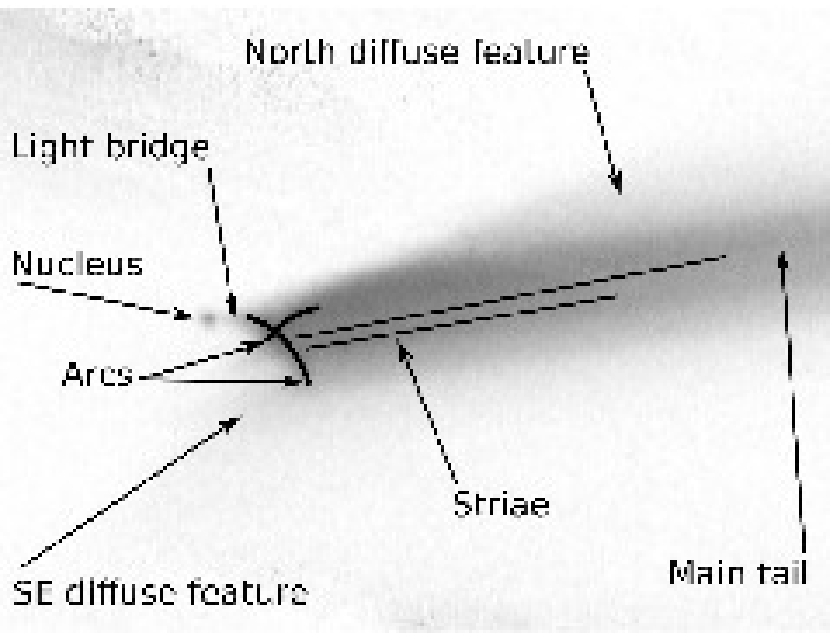}
\caption{Schematic of the main features of P/2010~A2. The background
  image is the 19~Jan. GN image from Fig.~\ref{fig:image}.e. See also
  Fig.~\ref{fig:enhance} for enhancement of the features.}
\label{fig:schematic}
\end{figure}

\begin{figure} 
\includegraphics[width=8.8cm]{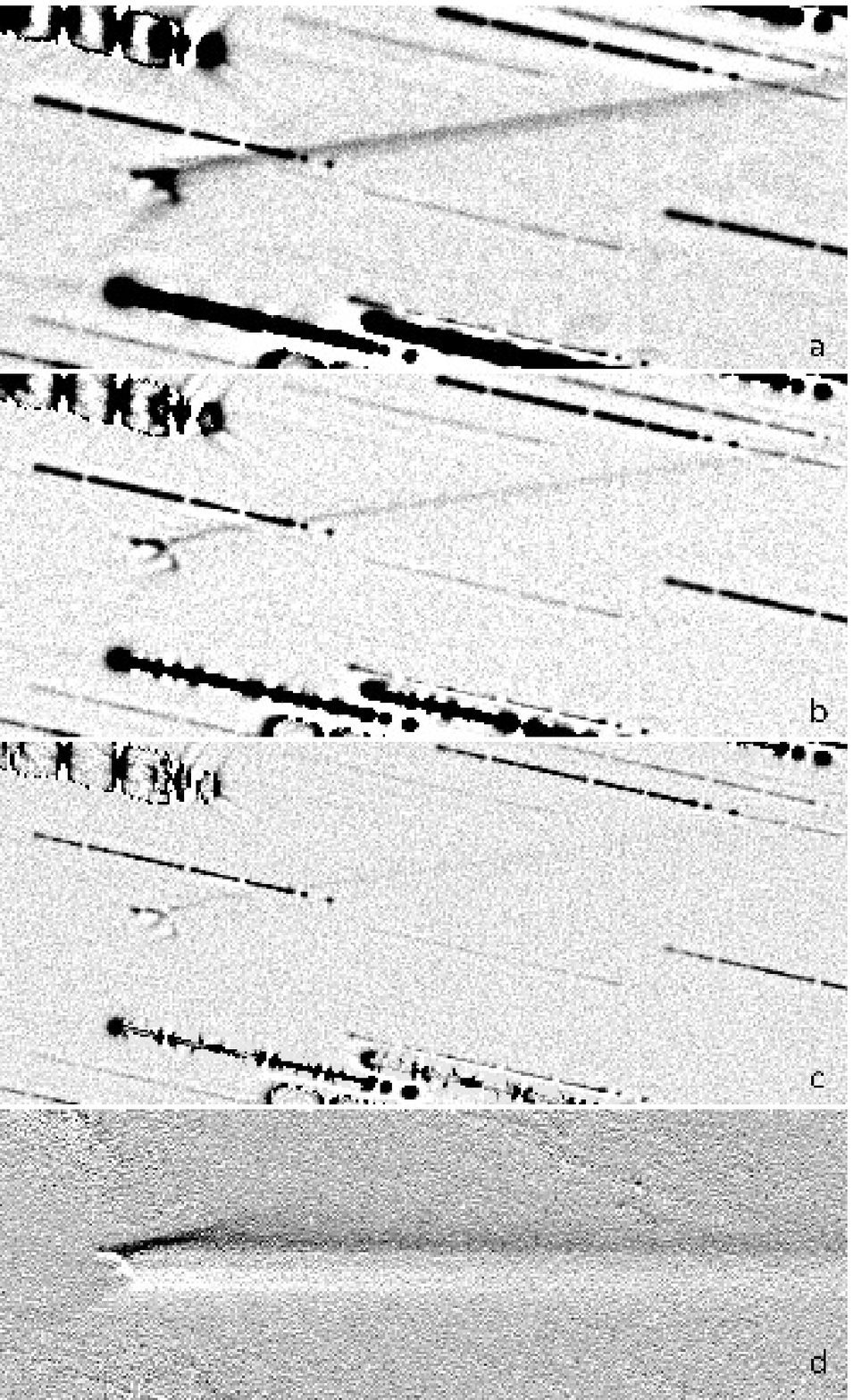}
\caption{ UT 19 Jan. 2010 GN image of P/2010~A2
  (Fig.~\ref{fig:image}.2), enhanced using unsharp masking, with a
  window of 20 (a), 10 (b) and 5 pixel radius (c). Panel (d) was
  produced by shifting the image by 1~pixel vertically and subtracting
  the result from the original, after rotating the image to align the
  tail along the rows.  
\label{fig:enhance}
}
\end{figure}


Following the nomenclature introduced by \citet{Jew+10} in their
Fig.~1 describing their Hubble observations, the main features of the
comet are a principal nucleus connected via a very faint and narrow
light bridge to an arc-shaped dust feature (see
Fig.~\ref{fig:schematic}). A second arc-shaped feature crosses the
first one almost perpendicularly. The main tail extends from these two
arcs. Very narrow striae along the main tail are emanating from knots
in the arcs. A low surface brightness fin-shaped structure lies above
(North) of the main tail. A second low surface brightness more diffuse
structure (not reported on the Hubble images) extends to the SE of the
arcs.

A variety of unsharp masking techniques were used to enhance the tail
structures.  A filter consisting of a sliding circular window of
radius $n$ pixels was moved through the 19 Jan. GN composite image and
the central pixel in the window was replaced by the median of the
pixels in the window.  The resulting smoothed image was subtracted
from the original image to remove the background thus enhance the high
frequency components.  The results for windows of 20, 10 and 5 pixel
radii are shown in Fig.~\ref{fig:enhance}.  The 20 pixel radius
filtering window brings out a broad dust feature at PA~$\sim
310\degr$, the ``North fin'', which is roughly in the anti-solar
direction. 
{The dust modelling described below gives some hints on the origin of
that features, which are discussed in Section~\ref{subsub:fin}.  
}
Additionally, the shifted individual GN
images were median combined, and rotated by 8.3$^{\circ}$ to place the
tail along a row.

The rotated image was then shifted by 1 pixel vertically and
subtracted from the composite to enhance structures along the dust
tail.  This is also shown in Fig.~\ref{fig:enhance}, panel [d].  Each
of the condensations of material in the arc are shown to be secondary
sources of dust.


The available data-set covers a period of 36 days, during which the
Earth crossed the object orbital plane (from -2.2$\degr$ to
0.6$\degr$). The morphology of the comet is remarkably constant over
that time, implying that the motion of the dust is slow, and that the
observed dust features have a significant thickness above the orbital
plane. This, in turn, implies the dust was ejected with a non-zero
velocity. We searched for evolution in the morphology between the
various epochs of observation, but given the observation interval and
resolution of the ground-based images, no changes in structure were
observed.

\subsection{The nucleus}

\subsubsection{Measurements}\label{sec:nucleus}


The nucleus brightness was measured on the two GN datasets, through a
series of apertures of increasing radius. The best compromise between
enclosed signal and sky background noise is found for an $0\farcs 3$
radius aperture. The nucleus magnitude measured in that aperture is corrected
for the missing flux using the star growth profile as a reference. The
resulting magnitudes are listed in Table~\ref{tab:photom}.

\begin{table*}
\caption{Nucleus photometry}     
\label{tab:photom}      
\begin{tabular}{llrrcrrrr}
\hline
\multicolumn{2}{c}{UT (exposure start)}&   Exp.time & Airmass & Filter & Seeing & Mag& M(1,1,$\alpha$)& M(1,1,0)\\
   &&   [s]      &         &        & [$''$] &  $r$' & $r$' & $r$'   \\    
\hline
\hline
2010-01-19 & 11:27:03.3 & 600. & 1.16 & r & 0.58 & 23.93$\pm$0.04 \\
2010-01-19 & 11:37:51.3 & 600. & 1.19 & r & 0.60 & 23.77$\pm$0.03 \\  
2010-01-19 & 11:48:38.2 & 600. & 1.22 & r & 0.63 & 23.96$\pm$0.04 \\  
2010-01-19 & 11:59:25.2 & 600. & 1.26 & r & 0.70 & 24.01$\pm$0.03 \\  
2010-01-19 & 12:10:12.1 & 600. & 1.30 & r & 0.81 & 24.20$\pm$0.04 \\  
2010-01-19 & average    &      &      & r &      &               & 22.34$\pm$0.04 & 21.74$\pm$0.04\\
\hline
2010-02-02 & 06:32:55.1 & 600. & 1.09 & r & 0.93 & *            \\
2010-02-02 & 06:43:42.9 & 600. & 1.08 & r & 0.93 & *             \\
2010-02-02 & 06:54:30.9 & 600. & 1.06 & r & 0.78 & 24.03$\pm$0.08 \\  
2010-02-02 & 07:05:17.8 & 600. & 1.05 & r & 0.92 & 24.38$\pm$0.06 \\
2010-02-02 & 07:16:04.8 & 600. & 1.03 & r & 0.75 & 24.00$\pm$0.06 \\
2010-02-02 & average    &      &      & r &      &               & 22.35$\pm$0.05 & 21.55$\pm$0.05\\
\hline
\end{tabular}

Note: *= the nucleus was contaminated by a field star.
\end{table*}



\begin{figure}
\includegraphics[width=8.8cm]{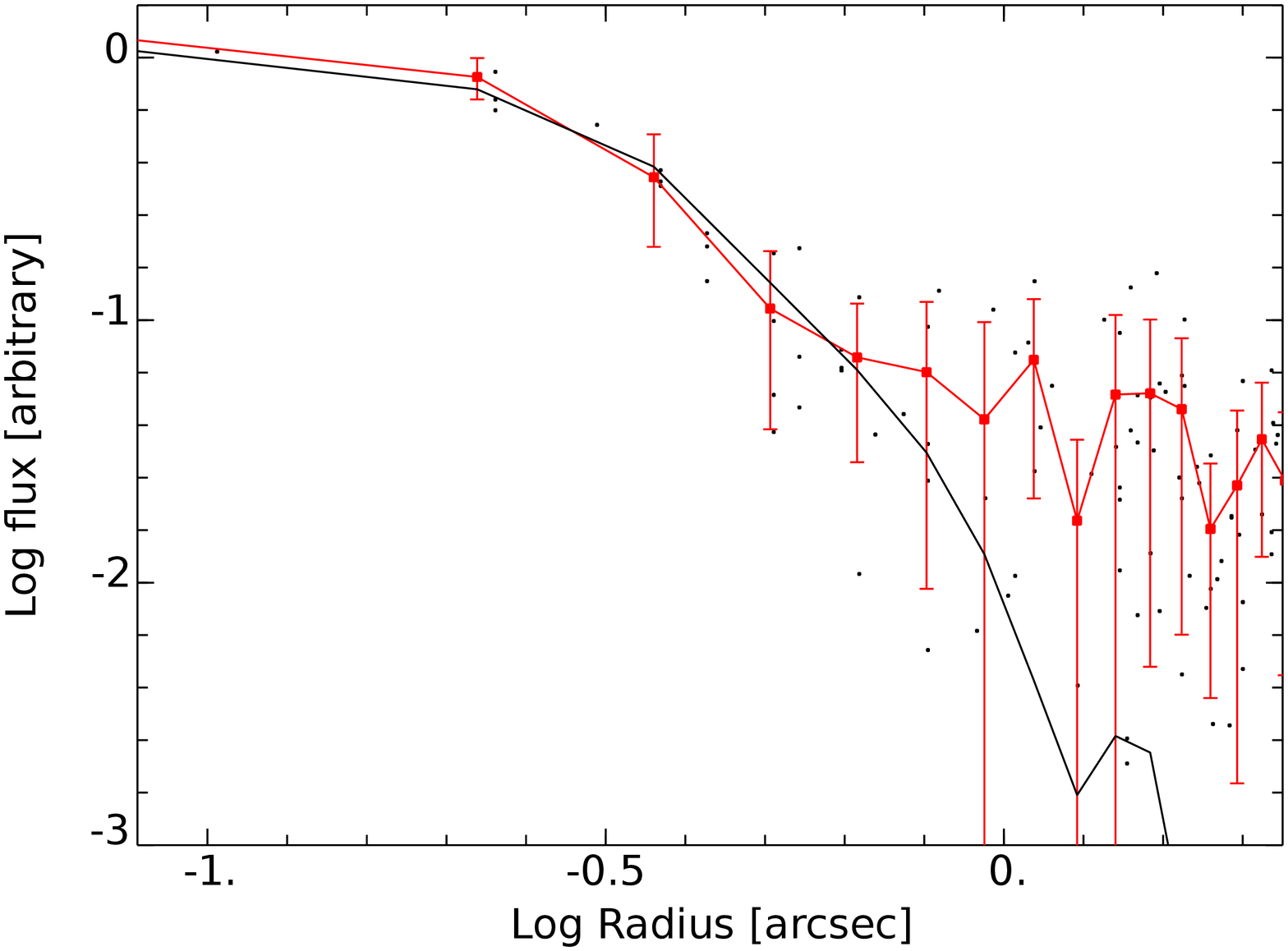}
\caption{Surface brightness profile of the nucleus between PA $-40$
  and 110$\degr$. The individual pixels are represented by small
  dots. The average brightness (between $-40$ and $110\degr$) in
  consecutive annuli is marked by connected solid disks. The other
  line is the scaled profile of a field star.  
}\label{fig:profile}
\end{figure}

The radial profile of the nucleus was measured between PA $-40$ and
110$\degr$ (i.e. the area not affected by the tail).  This was done by
converting the $(x,y)$ of each pixel to an angle and a distance, then
the flux is averaged between the two position angles for each distance
range. 
%
The surface brightness profiles are then constructed by iteratively
adjusting the asymptotic level of the brightness profile so that it
reaches zero.  The error bars shown in Fig.~\ref{fig:profile}
therefore reflect a robust estimate of the sky noise.  The average
profile is compared to that of a field star (measured perpendicularly
to the trailing). The profiles corresponding to the image with the
best seeing are displayed, normalized to the same peak brightness, in
Fig.~\ref{fig:profile}. The P/2010~A2 profile shows no significant
excess for radii smaller than 3 pixels ($0\farcs 42$), then a (noise)
extended contribution barely above the sky level.

In order to obtain a conservative estimate of the quantity of dust
directly surrounding the nucleus, the upper limit on flux excess from
the nucleus profile with respect to the reference star was integrated
out to a radius of $2''$.  The upper limit on the flux excess in an annulus is
defined as the measured surface brightness of the comet minus that of
the comparison star, plus a 1-$\sigma$ error bar, multiplied by the area
of the annulus. The limit on the flux excess integrated to a radius of
$2''$ corresponds to a mag $27.7$, or $<3$\% of the flux from the
nucleus. The nucleus did not present any significant mass-loss at the
time of the observations. This is further confirmed by the lack of any
dust visible in the lower-left (SE) quadrant below the nucleus, as
discussed in Section~\ref{subsub:duration}. This strengthens the
conclusion of \citet{Jew+10} that there was no cometary activity
based on the purely geometric variation of nucleus brightness with
heliocentric distances, withing 0.6~mag, from  25~Jan. until 29~May. 2010.

The average measured $r'$ magnitudes convert to absolute magnitudes
$r'(1,1,\alpha) = 22.34\pm0.04$ and $22.35\pm0.05$ for the two GN
runs. We do not have enough data to estimate the solar phase
effect. We therefore use the typical S-type asteroid $G=0.15$ to
correct for this effect, assuming the object is either a member of the
Flora family, or of the most frequent type in that region of the solar
system. The phase-corrected absolute magnitudes are $r'(1,1,0) =
21.74\pm0.04$ and $21.55\pm0.05$. The uncertainty is however now
dominated by the uncertainty of the phase correction, which could be
off by 0.08~mag if the object is asteroidal but not of the S-type
\citep{JLS02}, and by 0.10--0.15 if the object is a low albedo comet.
The magnitude difference between the two epoch is likely to dominated
by the rotational variability; indeed a body this small is likely to
have lightcurve amplitude of many tenth of magnitude.

Using the colour equations for the SDSS filters from
\cite{1996AJ....111.1748F}, the solar magnitude in the Sloan filter is
$r'_{\sun}=-26.95$ \citep{2001AJ....122.2749I}. Assuming a typical
S-type asteroid albedo of $p=0.11$, the absolute magnitude corresponds
then to a radius of 80 and 90~m 
for the two epochs. The formal uncertainty on
this value is completely dominated by the assumption of the albedo and
phase correction, and certainly includes a variation caused by the
rotation of the nucleus.
This is in agreement with the radius of 58--77~m reported by
\citet{Jew+10} using $p=0.15$ (which would convert to 68--90m with
$p=0.11$), and is marginally compatible with the $r=110 \pm 20$~m
from \citet{Mor+10}.

Assuming the body is spherical, with a density of 3\,000~kg~m$^{-3}$
---the average of 11~Parthenope and 20~Massalia, two
S-type asteroids \citep{Bri+02}--- its escape velocity is $v_e \sim
0.10$--0.12~m~s$^{-1}$.

If the object were instead a cometary nucleus (with a linear phase
correction of 0.04~mag/deg and an albedo $p=0.04$), then the radius
would be 120--140~m.
In this case, assuming a density of
1\,000~kg~m$^{-3}$, 
the escape velocity would be in the range $v_e \sim
0.09$--0.11~m~s$^{-1}$. As discussed later, we favour the asteroidal
nature of the object.

 Photometric measurements in the other filters have insufficient
signal-to-noise ratio to produce meaningful colours.


\subsubsection{Thermal model}\label{sec:thermal}

In order to examine the possibility of ice sublimation being the
driver of the observed dust ejection features, we treat the object as
a comet nucleus.  As such, the question is how long will ice survive
in the interior of such a small object at such a small heliocentric
distance.  For the purpose of modelling the thermal evolution of the
nucleus, we assume a similar 1-D model and numerical implementation to
that described in detail by \citet{1992ApJ...388..196P} and
\citet{2005PASP..117..796S}. Further discussion of the input physics
can be found in \citet{2004come.book..359P}.

A comet nucleus is generally portrayed as a porous aggregate of ices
and solids \citep{2004come.book...97W}. The internal structure can be
modeled as an agglomeration of grains made of water ice (and perhaps
other minor volatile compounds) and solids (silicates, minerals) at
some mixing ratio, with a wide spread in the size distribution of the
components.  The ice more volatile than water are neglected in our
treatment as any such ice becomes depleted during a long term
evolution, even down to the core of much larger objects
\citep[see][]{2009MNRAS.399L..79P}. Furthermore, if the object
originated in the asteroid belt, it is more likely that H$_2$O ice
would be crystalline and that the initial abundance of more volatile
ices would be minute \citep[e.g.][]{2004come.book..317M}. We solve the
set of coupled, time-dependent equations of mass conservation (dust,
ice and vapor components) and heat transfer \citep[see][for a detailed
  description]{2004come.book..359P}. The boundary conditions for these
evolution equations are vanishing heat and mass fluxes at the center,
vanishing gas pressures at the surface and a requirement for energy
balance at the surface:

\begin{eqnarray}
F(R) &=& \epsilon\sigma T(R,t)^4\,+\, \nonumber \\
& &     P_{\rm vap}(T)\sqrt{\frac{\mu}2\pi R_gT} H -(1-A){L_{\odot}
        \over 4\pi  r(t)^2}\cos \xi,
\end{eqnarray}
where $\epsilon$ and $A$ are the emissivity and surface albedo of the
nucleus, $L_{\odot}$ is the solar luminosity, $\sigma$ is the
Stefan-Boltzmann constant, $R_g$ the molar gas constants and $\mu$,
$P_{\rm vap}$ and $H$ are the molar mass, vapor pressure and sublimation
latent heat for water. This condition on the surface flux, $F(R)$,
depends on the Temperature $T$ , heliocentric distance $r$ and local
solar zenith angle $\xi$ and is calculated for each time step. The
first term corresponds to the re-radiated heat, the second to the
sublimation, and the third the absorbed solar radiation. Without
any additional information, we assume a fast rotator with the
sub-solar point illuminated at perihelion.

We consider two cases for this object: ``Case~I'', which is similar to
an S-type asteroid, and ``Case~II'', which is taken as a comet
nuclei. Using the assumptions and measurements described in the
previous section, Case~I has a radius, bulk density and albedo of
85~m, 3000~kg~m$^{-3}$ and 0.11, respectively. Case~II has a radius,
bulk density and albedo of 130~m, 1000~kg~m$^{-3}$ and 0.04,
respectively. If we assume the solid component (both the nucleus and
the dust) to have specific density similar to that found for the dust
particles of comet Wild 2, $\sim$ 3400~kg~m$^{-3}$
\citep{2009M&PS...44.1489K}, we can constrain the initial ice
abundances for the thermal evolution calculations
\citep[see][]{2008SSRv..138..147P}. These values are taken as 0.05 and
0.5 (with corresponding initial porosities of 0.01 and 0.3) for Case~I
and Case~II, respectively. All other physical parameters are taken to
be similar to the \citet{2009MNRAS.399L..79P}.

\begin{figure}
{\bf a.} \includegraphics[width=8.cm]{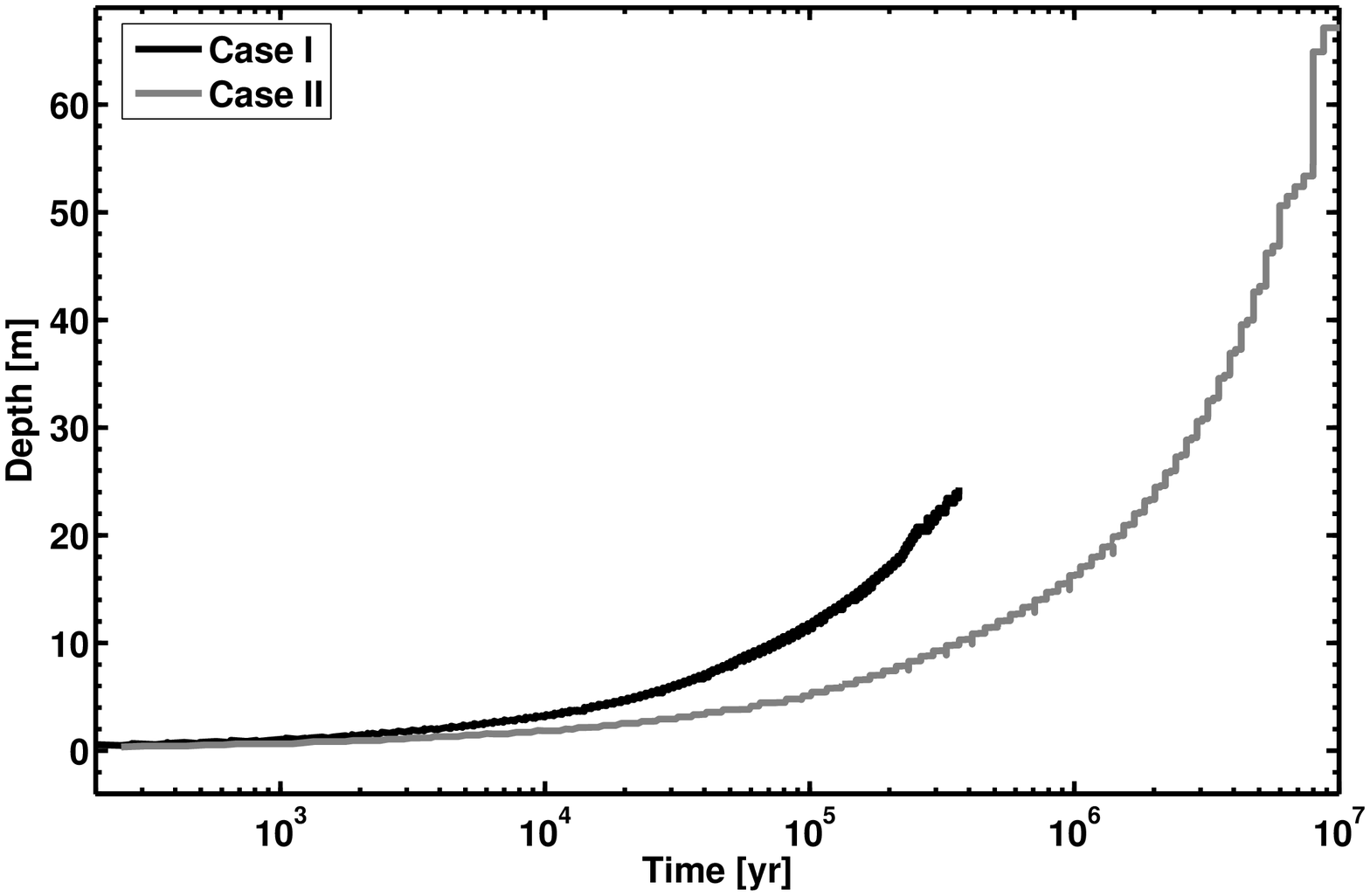}\\
{\bf b.} \includegraphics[width=8.cm]{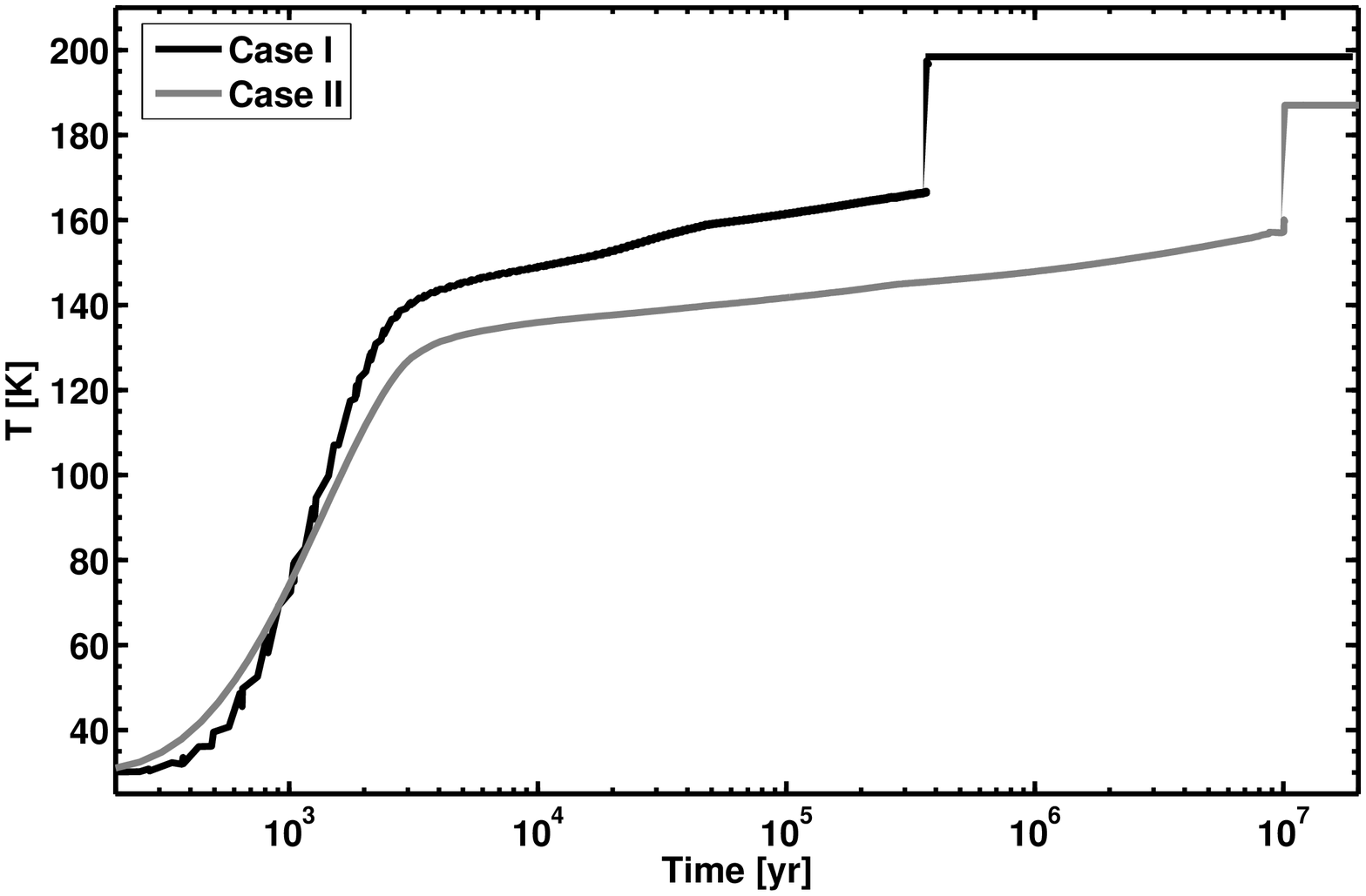}
   \caption{Evolution of the water ice sublimation front and the central temperature for the nucleus of
     P/2010~A2. The x-axis represents the time [yr] in log scale since the placement of the object in its current orbit, the y-axes
     represent: {\bf a.} The sub-surface depth where crystalline ice survives; {\bf b.} The central temperature. Case~I (``asteroid'') is plotted in black, while Case~II (``comet'') is plotted in gray.
}
     \label{fig:thermal1}%
\end{figure}

Our simulations of the nucleus evolution are run until the water ice
component is depleted from the interior. For Case~I (``asteroid''),
there is no more water ice present after $\sim 3\times 10^5$~yr, while
for Case~II (``comet'') water ice becomes negligible and is buried
deep under the surface after several $10^6$~yr. This is shown in
Fig.~\ref{fig:thermal1}a, where the sub-surface
water ice depth is plotted as a function of time. 
Beyond this point in time, the water ice component is rapidly
depleted, as the innermost layers reach a temperature of 200~K and
180~K, for Case~I and Case~II respectively. This is shown in
Fig.~\ref{fig:thermal1}b, where the evolution
of the central temperature is plotted, for both cases, up to
$10^7$~yr. The penetration of heat into the interior and the rise of
the innermost temperature is slower for Case~II and the final
temperature reached is lower throughout most of the evolution
time. Case~I heats up faster because it is less porous (bulk density
is higher) and has a much larger mass fraction of dust (silicates and
minerals), so heat diffusion is less attenuated by sublimation.

In summary, considering a comet-like and an asteroid-like cases for
the composition and structure of the nucleus, our thermal modelling
efforts indicate that the interior reaches a temperature over 180K in
a few $10^5$ to a few $10^6$ years. The nucleus would therefore be
fully depleted of water ice ---and any other more volatile ice--- in
that time scale. The orbit of P/2010~A2 is typical of an inner Main
Belt asteroid, possibly member of the Flora family, with no indication
that it would have been very recently injected or captured from a
location more distant from the Sun. The survival depth of water ice
becomes greater than $\sim 10\%$ of the object's radius in less than a
million years, almost independent of the initial choice of model
parameters. This model therefore indicates that ice sublimation could
not play any role in lifting dust from the surface of the object or
dragging any substantial amount of grains from the interior. In what
follows, we will therefore assume that the nucleus of P/2010~A2 has
the characteristics of a Flora family, S-type asteroid, as detailed in
this section and the previous one.


\subsection{Dust and the tail}

\subsubsection{Dust dynamical modelling}
\label{sec:FP}


The dust tail was modeled using the Finson-Probstein (FP) dust
dynamical method \citep{FP68}, modified by \citet{farnham96}.

Dust is extracted from the nucleus, by gas drag in traditional
sublimation-driven cometary activity, or via direct ejection in the case
of an impact.  The dust is decoupled from the gas flow (if any) within
a few nuclear radii, and is no longer influenced by the nucleus
gravity, which becomes negligible.  The motion of a refractory dust
grain will then be defined by its initial velocity and affected by the
solar gravity and radiation pressure, both acting in opposite radial
directions. The net result of the two influences may be thought of as
a reduced gravitational force acting upon the dust. The parameter
$\beta$ is defined as the ratio of the radiation pressure force to the
gravitational force, and is given by
\begin{equation}
\beta = 5.740 \times 10^{-4} \frac{ Q_{\rm pr} }{\rho
  a} 
\label{eq:beta}
\end{equation}
for grain of radius $a$ [m] and density, $\rho$=3\,000~kg~m$^{-3}$.
$Q_{\rm pr}$ is the radiation pressure efficiency (typically 1--2,
depending on the material scattering properties).

The FP method involves computing the trajectories of dust grains
ejected from the nucleus at velocity $v$ and under the influence of
solar gravity and radiation pressure.  A collection of particles of
different $\beta$ ejected at a particular time follow trajectories
called synchrones, while a single particle size ($\beta$) ejected over
a range of times follow trajectories called syndynes.  The combined
set of curves maps out the expected ejected dust distribution.  The
extent to which the actual dust surface brightness correlates with the
syn-curves may be used to infer which grain sizes are present, the
onset an duration of activity, and any initial ejection velocity.

In order to investigate the origin of the dust tail, rather than
starting with an entirely generic model which would have a large
number of loosely constrained free parameters, we proceeded by steps,
focusing on some aspects of the tail and making some simplifying
hypotheses.

\begin{figure*}
   \includegraphics[width=18cm]{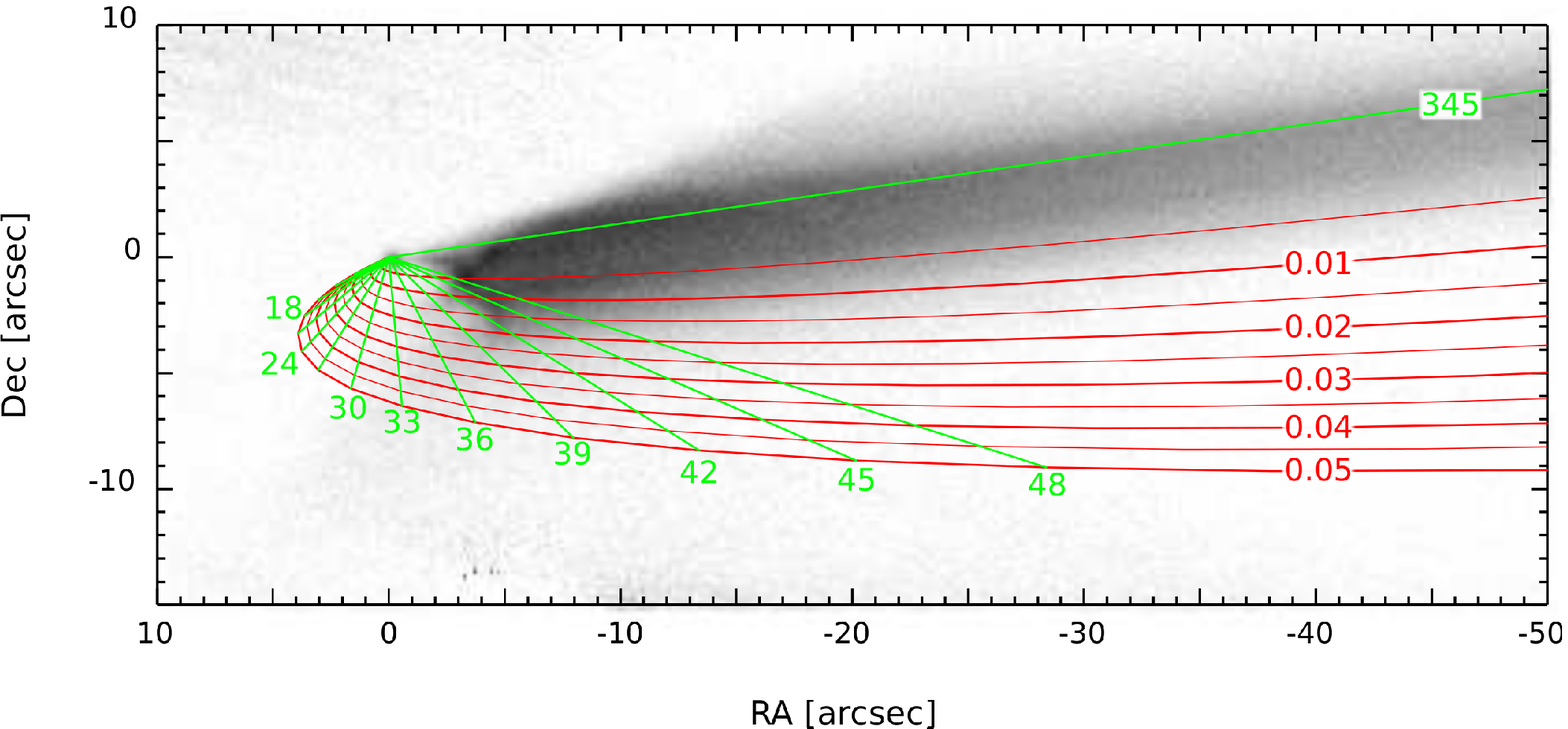}
   \caption{Syndynes (labeled with $\beta=0.01$--$0.05$) and
     synchrones (labeled in days since emission, 18--345) for dust grains
     emitted from the nucleus with zero velocity. The background image
     if from Fig.\ref{fig:image}.e.
}
              \label{fig:fp}
\end{figure*}

\subsubsection{Tail morphology and duration of the activity}

\label{subsub:duration}
\label{subsub:overall}

Overlaying the syndyne and synchrone curves onto the image of the comet
(see Fig.~\ref{fig:fp}), it is clear that the region corresponding to
dust emission during the 30~days before the observations is completely
empty of dust. Together with the negligible contribution from the dust
to the near-nucleus area, this confirms that any cometary activity
must have stopped at least several weeks before the observations.

The light bridge connecting the nucleus to the tail matches the
position expected for very large dust grains. \citet{Jew+10} measured
the orientation of the tail on the high-resolution HST images,
concluding that there was a short period of dust emission between
Feb. and Mar. 2009.

However, while the orientation of the tail corresponds well to the
direction of that synchrone, its position is offset from what is
expected from a dust emission a zero speed from the nucleus.  Also,
the position of the northern fin is not matched by any syndyne, and
the constant appearance of the tail while the Earth passed through the
orbital plane indicates a more complex, 3D geometry and dust emission
with non-zero velocities.

We will now further investigate the dust release
characteristics. Adding up the scattered light from a distribution of
dust particles emitted from the nucleus over a range of times as it
moves away from the object, one can build synthetic images of the
corresponding tail. Comparing the surface brightness of the synthetic
tail to the real data, observational data can be inverted and
information about the grain size distribution, ejection velocity and
onset and cessation of activity can be obtained.


Various general dust emission models were explored to investigate the
overall morphology of the tail.  The first model included emission for a
long duration of time, about a year from the time of observation to three
months before the observation. 
The second model included only one month of emission beginning a year
prior to the observation.
Finally, a third model only included a single burst of emission one
year before the observation. The emission takes place at the sub-solar
point, with a constant dust production during the considered period.
The dust size range and distribution and emission velocities 
were adjusted so that the reconstructed tail reproduces as well as
possible the observations. The parameters are adjusted to reproduce
the orientation (measured by its position angle) and width (measured
perpendicular to the tail) of the tail, and the  surface-brightness
distribution along the tail. The actual position offset between the
tail and the nucleus is not measured at this stage; it is considered
as an effect of the global/average emission velocity, and will be
dealt with later.

\begin{figure*}
 \sidecaption
{\bf a.} \includegraphics[width=12cm]{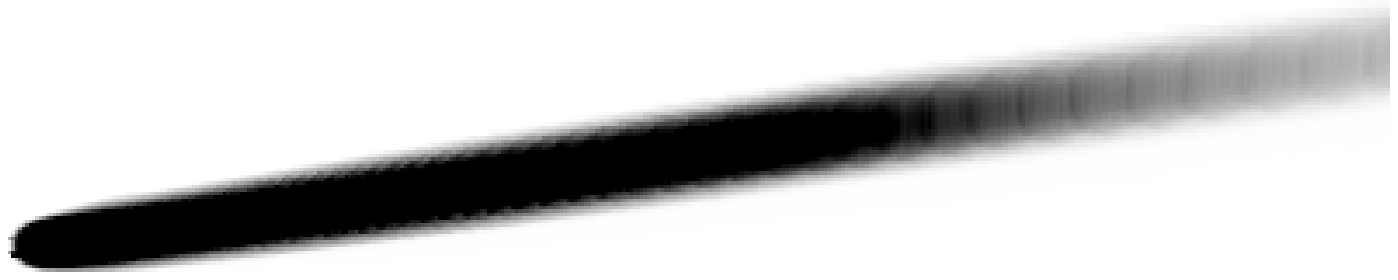}
   \caption{~ Synthetic image of P/2010~A2, {\bf a.} for the best fit
     with a long period of dust emission (Model~1, in
     Table~\ref{tab:fp}). {\bf b.} for the best fit with a month-long
     period of dust emission (Model~2). {\bf c.} for the best fit
     with single burst of dust emission (Model~3). {\bf d.} Reference
     image, same scales. This is the same image as Fig.~2.e.
      }
{\bf b.} \includegraphics[width=12cm]{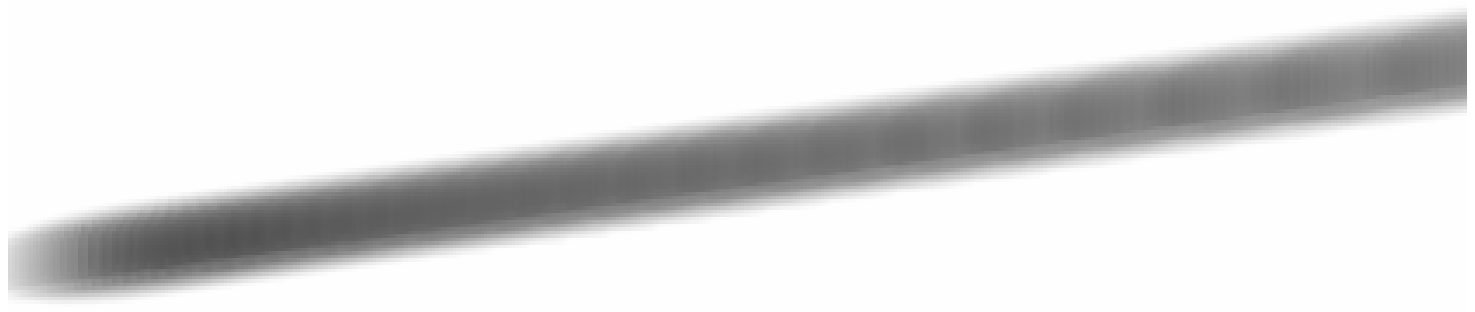}\\
{\bf c.} \includegraphics[width=12cm]{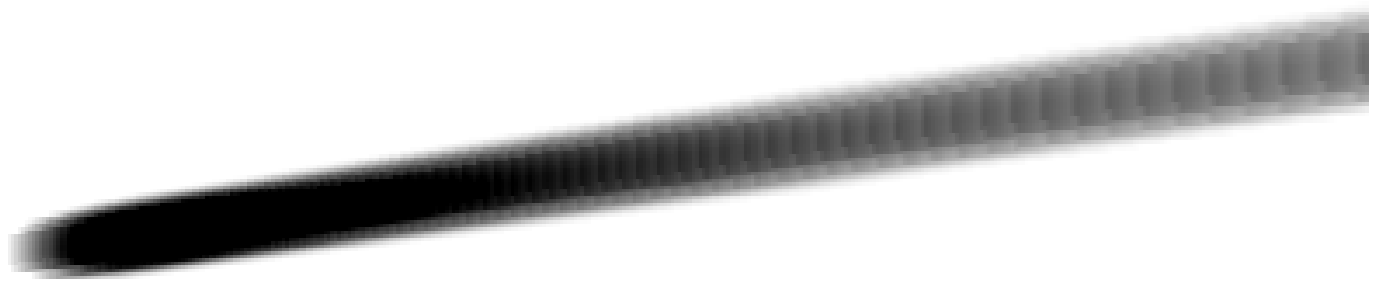}\\
{\bf d.} \includegraphics[width=12cm]{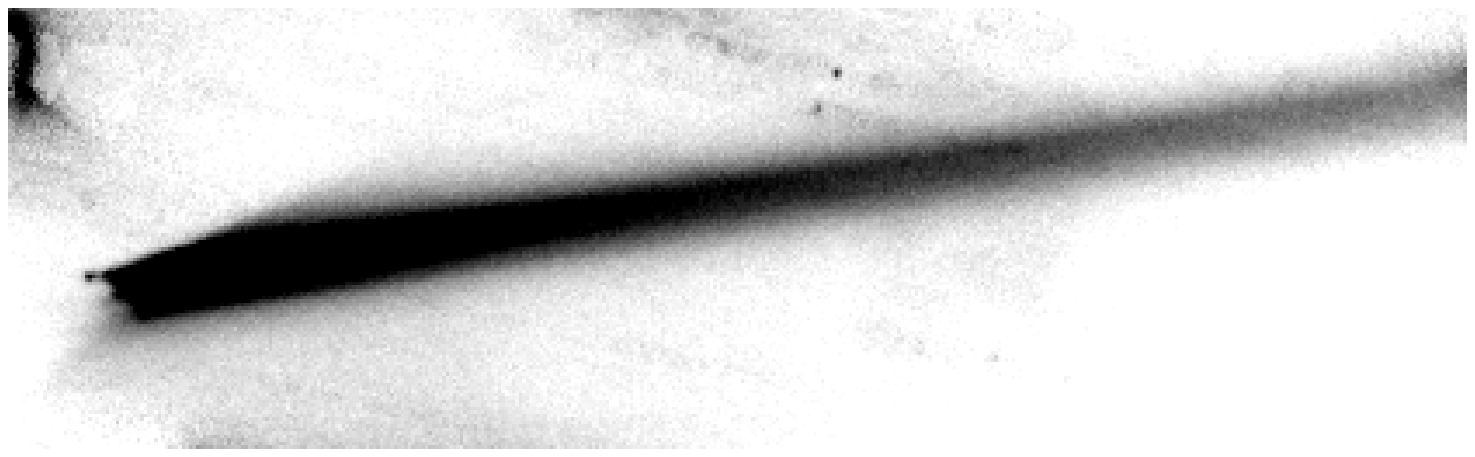}
    \label{fig:fpTony}
\end{figure*}

\begin{figure}
 \includegraphics[width=8.8cm]{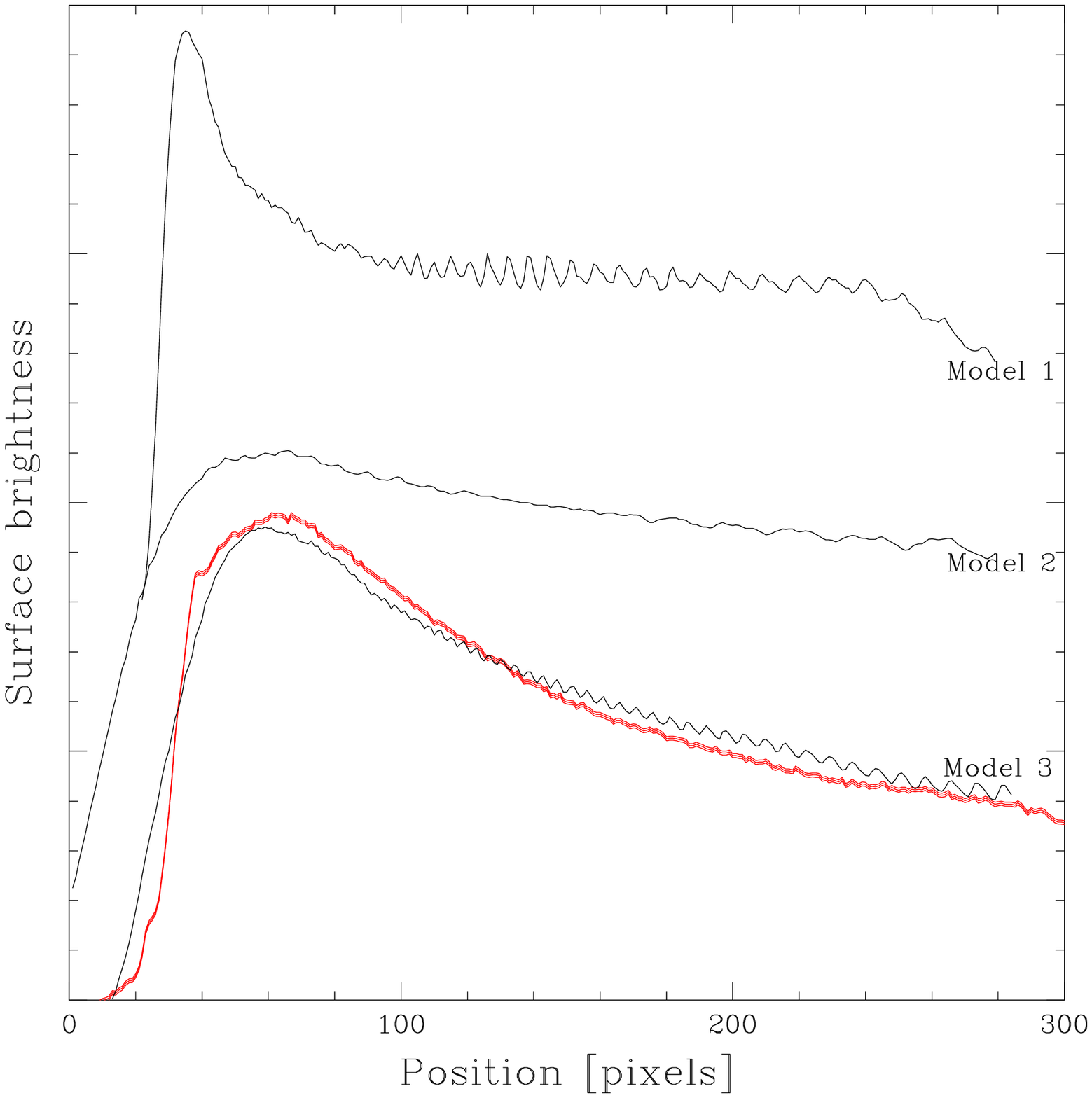}
 \caption{
Photometric profile of the reference image (bold line) and the three
models. Model 1 was shifted vertically for clarity.
}
\label{fig:fpProf}
\end{figure}

The best parameters for each model are listed in Table~\ref{tab:fp},
the corresponding reconstructed images are shown in
Fig.~\ref{fig:fpTony}
{and the surface brightness profiles are displayed in
  Fig.~\ref{fig:fpProf}. These profiles were obtained by averaging 52
  pixels perpendicularly to the tail on an image rotated by 8$\degr$
  so to have the tail along the X direction. }
The first two models failed to reproduce the
overall width, shape and/or brightness profile of the tail, even for
the best adjustment of the parameters. While it may not be obvious
from Fig.~\ref{fig:fpTony}, the light distribution along the length of
the tail is very different for models 1 and 2 compared to the
reference image, 
{as seen in Fig.~\ref{fig:fpProf}.}

The third model, with a single burst of emission, reproduces well
these shape, width and profile of the tail with the particle sizes in
the $a=1.5$ to 20~mm range, with a size distribution proportional to
$a^{-3.5}$.  The low end of the range is set by the edge of
the image, and the upper end corresponds to particle with a minimal
sensitivity to the radiation pressure.

The parameter values for Model 3 are tightly constrained in the
adjustment process: power laws shallower (-3.75) or steeper (-3.25)
fail to reproduce the profile of the tail.  The best range of
speeds of the particles was 0.20 to 0.30~m~s$^{-1}$, with smaller particles
having higher velocities.  The range of particle speeds that still
develop an acceptable model is only 0.10--0.40~m~s$^{-1}$. The total emission
can vary by less than a factor of 2 and still acceptably reproduce the
observed surface brightness.  The total amount of mass ejected
(assuming a density $\rho=$ 3\,000~kg~m$^{-3}$ and albedo $p=0.11$)
was about $5\times 10^{8}$ kg, again
within a factor of 2.

In summary, reproducing the overall morphology of the tail suggests
that the dust was emitted during a single burst about one year before
the observations. The dust grain distribution follows a power law in
-3.5 in the 1.5--20mm range, emitted with slow velocities in the
0.20--0.30~m~s$^{-1}$ range.

We also note that, for a water-ice sublimation-driven dust emission,
\citet{Del82} obtains for an object at $r\sim 2$~AU ejection
velocities in the 0.3--0.4~km~s$^{-1}$ range, i.e. over 3 orders of
magnitude larger than the velocities obtained in the FP
modelling. However, the assumptions made by Delsemme are not
necessarily valid here, so we will not use this argument.

\citet{Mor+10} analyzed a set of images obtained between 14 and
23~Jan.~2010 using a different modelling technique, assuming particles
ranging between 0.01 and 10~mm with a density of 1000~kg~m$^{-3}$ and
an albedo of 0.04. Their best fit suggests dust emission lasted over
several months, ending just after perihelion (Dec.~2009), with a
complex emission function over that period.  While their model was
richer than ours in terms of the complexity of the emission pattern,
our first FP analysis cannot support that there was dust emission that
late (see Fig.~\ref{fig:fp}). Our model that is closest to their best
fit is Model~1 (although ours uses a constant emission). Our Model~3,
with an burst emission, reproduces the tail much
better. \citet{Jew+10} concluded from the evolution of the tail
position angle over 25~Jan.--29~May. 2010 that the dust had been
emitted in a brief burst around 2~Mar.2009, plus or minus a few
weeks. \citet{Sno+10}, taking advantage of the different viewing
geometry from the Rosetta spacecraft, concur that the emission was
very short, further constraining the emission date to 10~Feb. 2009,
plus or minus 5 days.

Model~3 nicely reproduces the overall morphology of the tail, in
particular its orientation and width, and the detachment from the
nucleus and the overall surface brightness profile. However, it fails
to reproduce the sharp edges and features visible in the head. Using
the results from this model, we will further investigate the
hypothesis of a single burst of emission.

\begin{table}
\caption{Three FP models}
\label{tab:fp}
\begin{tabular}{p{4cm}lll}
\hline                           
Model                             &1        & 2         &3      \\
\hline                           
Emission period,                  &         &           &       \\
starting on 10 Feb. 2009          &250 days &30 days    &1 day  \\
\hline                            
\hline                           
Particle size range (mm), $a$     &0.3--3   &1.5--100   &1.5--20\\
Particle size distribution power  &-1       &-3.9       &-3.5   \\
Emission velocities   (m/s)      &0.20-0.30 &0.10-0.20  &0.20-0.30  \\
\hline                                                    
Fidelity of the model             &Poor     &Very poor  &OK     \\
\hline
\end{tabular}
Note: the emission duration is set, the other
  parameters are adjusted to reproduced the observed tail.
\end{table}



\subsubsection{Quantity of dust in the tail}\label{sec:dust}

The presence of striae in the tail is usually a sign that secondary
dust emission is taking place, for instance from sublimating grains in
traditional comets. It is also a warning that basic FP modelling
should be applied with extreme care. In the case of P/2010~A2,
however, the situation is actually simpler. Considering the bright
knots in the arcs as the secondary dust grain source giving birth to
the corresponding striae,
{the straight striae emanating from the knots
follow synchrone curves. 
}
The range of emission epoch compatible
with the observations is measured esitimating (by eye) the range of
position angle that fit that of the striae. This range corresponds to
emission taking place $340^{+30}_{-20}$ days before the observations,
i.e. in agreement with the epoch of the main event. We conclude that
each stria corresponds to dust grains with a distribution of $\beta$,
ejected from the nucleus at the same time of the main dust release,
with the same velocity as the material in the knot leading the
stria. The surface brightness distribution as a function of the
distance to the arc constitutes therefore a ``dust grain radius
spectrum'' of the material.

We wish to estimate that dust size distribution, as well as the total
quantity of dust in the tail, using the sky-subtracted, flux
calibrated composite image from 19 Jan. 2010 (the deepest image).  For
each pixel we want to know which fraction is covered by dust (filling
factor $f$), multiplied by the albedo ($A$).  $Af$ is a convenient
dimension-less equivalent to the surface brightness.  Using the
formalism of \citet{AHe+84} we convert the measured flux in $Af$, ie.
\begin{equation} \label{eq:AH}
Af = \left(\frac{2 \Delta_{\rm km} r_{\rm AU}}{\rho_{\rm km}}\right)^2
   \frac{F_{\comet}}{F_{\sun}}   ~, 
\end{equation}
$\Delta_{\rm km}$ is the geocentric distance in km, $r_{\rm AU}$ the
heliocentric distance in AU, $\rho_{\rm km}$ the linear size of the
aperture at the distance of the comet (in km), $F_\comet$ the measured
flux of the comet, and $F_{\sun}$ the flux of the Sun at 1AU.
Converting Eq.~\ref{eq:AH} for square pixels ($p$ arcsec), and
expressing ${F_\comet}/{F_{\sun}}$ in CCD ADUs, it becomes
\begin{equation}
Af = \frac{4 \pi r^2   }{(4.848\times10^{-6} p)^2} 
   10^{-0.4 (ZP -k(z-1)-M_{\sun} )} \,  {\rm ADU}_\comet ,
\end{equation}
where $ZP$ is the zero point, $z$ the airmass and $k$ the extinction
coefficient (in mag/airmass; Gemini's ZP is expressed for 1 airmass,
$ZP=28.20$ from Section~\ref{sec:dataproc}).

\begin{figure}
    \includegraphics[width=8.8cm]{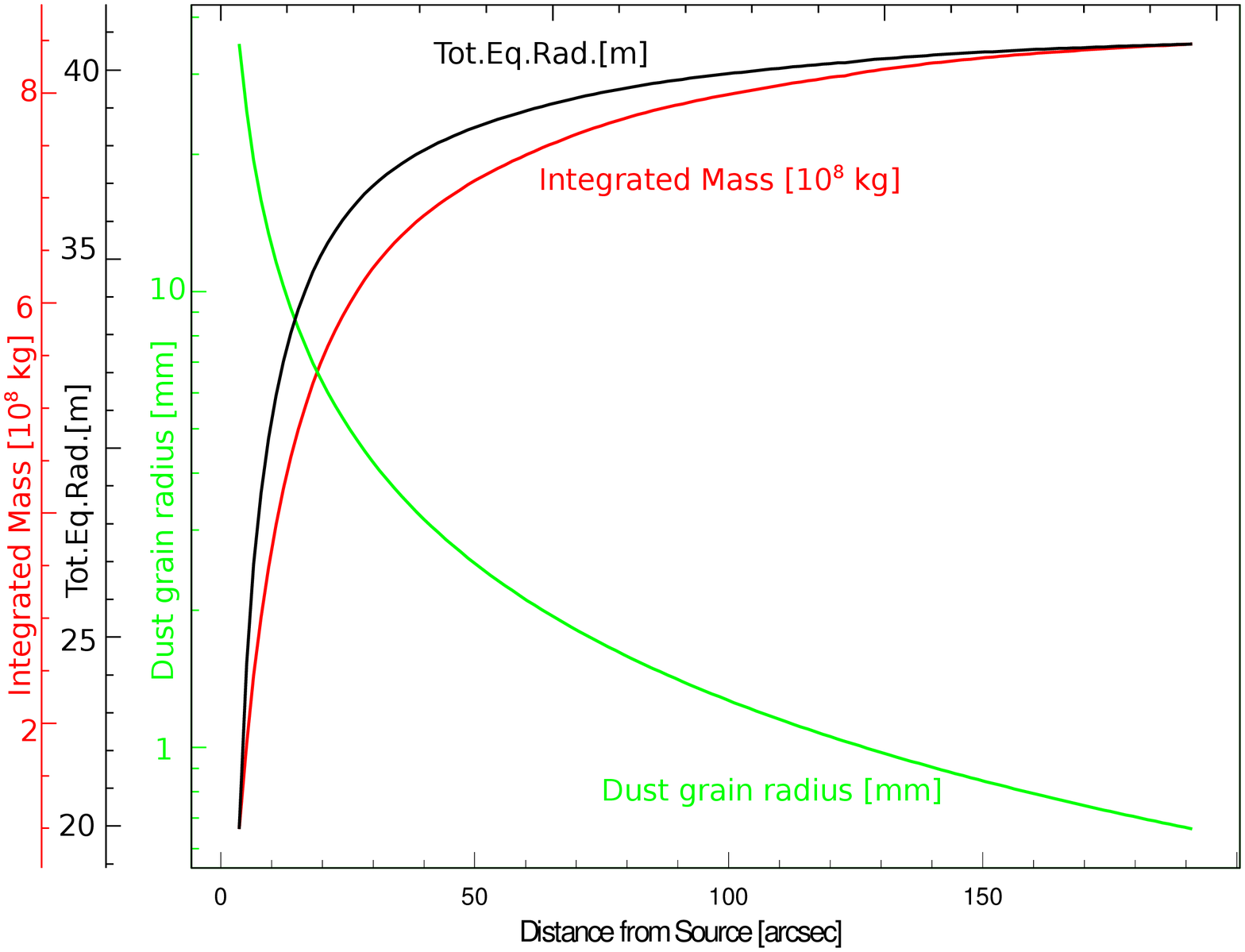}
    \caption{The labeled curves show, as a function of the distance to
      the arc-shaped features in the head of the tail, the radius of
      the dust particles [mm], the integrated mass in the tail out to
      that distance [kg] and the radius of the sphere of equivalent
      volume [m]. Refer to the text for the assumptions that were used
      to derive these values.}
    \label{fig:chunk}
 \end{figure}

In order to count the number of dust particles in each pixel, we
estimated the size of the particles, assuming that all the dust in
the tail was emitted during a single burst that took place on 
10~Feb.~2009 (using the date of \citet{Sno+10}, see
\ref{subsub:duration}), with a zero velocity from the large chunks of
material that 
ended up located in the arcs. The apparent distance from the arcs to
the particle is then a measurement of $\beta$, the conversion factor
is estimated running our FP code for particles over a range of values of
$\beta$ for an emission date in Feb.~2009. Using the relation between
$\beta$ and the characteristics of the particle from \citet{FP68}, we
convert $\beta$ in the radius of the particles, $a$, using
Equation~\ref{eq:beta}.  We use $Q_{\rm pr}=1$, as \citet{M09}, for
large absorbing grains \citep{BLS79}. We assume a density of
3\,000~kg~m$^{-3}$ (see Section~\ref{sec:nucleus}).
Figure~\ref{fig:chunk} displays the corresponding sizes, ranging from
few tenths of millimetres at the edge of the frame to few millimetres
close to the arcs. Using the formalism of \citet{Aga+07}, the dust
size distribution inferred from the surface brightness profile along
the tail is a power law in the $a=1$--20~mm range, with a number of
particles proportional to $a^n$ 
{(differential size distribution)}, 
with $n={-3.44\pm0.08}$. This value is in agreement with that modeled
in the previous section, and with those reported by \citet{Mor+10},
$n=-3.4\pm 0.3$, \citet{Jew+10}, $n=-3.3\pm 0.2$
{and \citet{Sno+10}},
$n=-3.5$. A value close to -3.5 is significant because this is in
agreement with the empirical value for the distribution of particle
sizes resulting from a collision, and for a collisionally relaxed
population \citep{Doh69}. The value for the dust in the tail of
P/2010~A2 is slightly steeper than the size distribution measured with
Hayabusa for the boulders larger than 5~m on the surface of
(25143)~Itokawa, $n=-3.1$ \citep{Nak+08}.

The total $Af$ is measured in a series of thin rectangular apertures as
long as the width of the tail and covering it from the arcs until the
edge of the frame. The total $Af$ is converted into a number of
particles using the radius $a$ of the particle at that distance from
the arcs as computed above. 

The total mass and volume of dust is integrated by adding the
contribution of each rectangular aperture starting from the arcs. The
mass integrated along the tail, as well as the volume of the
corresponding sphere are displayed in Fig.~\ref{fig:chunk}.While there
is still a significant amount of dust in the tail extending beyond the
edge of the frame, the asymptotic behaviour of the growth curve
indicates that a very large fraction of the mass is present in the
visible part of the tail. With these assumptions, we estimate that the
tail contains at least $8\times 10^8$kg of ejected dust, which could be
re-assembled into a sphere of radius $r_e \sim 40$~m. 


\citet{Jew+10} reached similar estimates based on the analysis of
their Hubble images: 6--60$\times 10^7$~kg, corresponding to a sphere
with a radius of at least 17--36~m. The latter corresponds to a
distance of 20$''$ from the arcs on Fig.~\ref{fig:chunk}, which is
about where the S/N starts to degrade on their Hubble images: we
consider their larger estimate in agreement with ours.  The total
ejected mass is likely to be higher, as these values poorly estimate
the size and number of large particles in X-shaped arcs.

\subsubsection{Envelope of the tail}
\label{subsub:envelope}

A dust emission model was developed to investigate the possible cause
of the position of the tail behind the X-shaped arcs, and its distinct
detached geometry. For this investigation, rather than
trying to reproduce the light distribution in detail using a
general fit of the dust velocity vector ${\bf v}_{\rm d}$ and $\beta$
distributions, a greatly simplified model was implemented as follows. The
details of the model, of its implementation and of its results are
beyond the scope of this paper, and constitute a separate, companion
paper \citep{PAPER2}. 

The dust emission is represented by an instantaneous burst on 10
Feb. 2009 \citep[date from][see \ref{subsub:duration}]{Sno+10}. The first
simplification represents the emission geometry with a cone
(defined by its half-opening angle $\alpha_{\rm c}$ and axis orientation),
on which dust grains characterized by a distribution of $\beta$ are
emitted with a distribution of velocity $v_{d}$. The motion of the
dust grains is computed taking into account the Sun's gravity and the
radiation pressure to determine their final positions in the image
plane at the time of the observations.

The second simplification is to only consider whether the dust grains
fill the envelope of the observed tail, completely neglecting the
light distribution within that envelope. The envelope is defined
visually on the images of 19~Jan. (Fig.\ref{fig:image}.e), so as to
enclose the main tail. Several different plausible envelopes,
estimated by eye, were defined to
test the dependency of the result on the choice of the envelope.  This
allows us to neglect the actual distributions of $\beta$ and $v_{\rm
  d}$: only the minimum and maximum values are relevant to the
model. The natural minimum velocity and minimum dust size are zero;
the dust size is assumed to be independent of the velocity
{(this is
supported by experimental studies of impact onto porous, regolith-like
material, which are appropriate for gravity- or weak cohesive
strength-controlled craters which, as we discuss later, seems to be
the case here). 
}
Finally,the only difference between a filled emission cone and a
hollow one is the presence of a cap at the open end of the cone, we
can therefore further simplify the model considering a hollow cone.

This leaves only four free parameters: the two coordinates defining
the orientation of the cone axis, the cone half-opening $\alpha$, and
the maximum dust velocity $v_{\rm d,max}$. The merit function of a set
of parameters is computed by assigning penalty to dust grains ending
up outside of the envelope of the tail, and to pixels within the tail
that contain no dust grain. \cite{PAPER2} describes the exploration of
the four parameter space, the method used and the sensitivity to the
definition of the envelope. Only a fairly narrow subset of the
parameter space can reproduce the observed tail envelope. The best
model is a cone (either full or hollow) with a half-opening of
$40\degr$, pointing below the orbital plane and on the forward side of
the comet, with a maximum dust velocity of 0.2~m~s$^{-1}$. The
direction of the emission is strongly constrained by the position of
the tail with respect the nucleus and by the sharp edge of the head.

\subsubsection{Arc-shaped features}
\label{subsub:arc}

In order to investigate the origin of the arc-shaped features (see
Fig.~\ref{fig:schematic}), another model was implemented; it is also
described in detail in \cite{PAPER2}. Each arc appears as a
one-dimensional feature; they are not parallel to the synchrones for
the date of the dust emission (see Fig.~\ref{fig:fp}, position angle
near $278\degr$). They can therefore not be caused by the radiation
pressure-driven spread of particles with a range of $\beta$; they must
therefore correspond to very large particles very weakly sensitive to
radiation pressure (i.e. with $\beta =0$).

A large number of particles were modeled, with an emission date on
10~Feb. 2009 \citep[date from][see \ref{subsub:duration}]{Sno+10},
sampling the emission direction and velocity space. The motion of
these particles was then computed until the date of the observations,
and their position in the image plane computed. Those landing on
position of the arcs are marked. Only particles with a ejection
velocity $v_{\rm d} > 0.20$~m~s$^{-1}$ can contribute to the X-shaped
arcs, but this model cannot set an upper limit. Indeed, dust grains
with higher emission velocity can be emitted and end up further away
from the nucleus at the time of the observation, but still appearing
in the same position in the image due to foreshortening. Test particles
from two regions in the (velocity, direction) space end up in the
position of the arcs. The arcs could therefore have been produced by
dust grains from one of these region, the other, or a combination of
both. 

However, one of these regions is found to match a section of the
surface of the cone modeled to describe the overall tail in the
previous section: the direction of emission and velocity are
identical. This suggests that the same hollow cone (with its orientation,
opening and emission velocity) describes both the overall tail geometry
as well as the arcs in the head of the tail. The former corresponding
to small dust grains subjected to radiation pressure, the latter
corresponding to a fraction of the large dust grains populating the
propagated cone. Only a fraction of the large dust grains are visible
as arcs, as the geometry at the time of the observations gives a
foreshortened side view of a segment of the cone. The other large dust
grains appear diluted over the tail.

Observations from another point of view should show other features
fortuitously aligned along that other line of sight. Unfortunately,
the Rosetta images \citep{Sno+10} do not have a sufficiently high
resolution to reveal them: 
{all the region presenting structure was
contained in one single pixel of the Osiris Narrow Angle Camera on
Rosetta. }
In this context, the striae visible in the tail (see
Fig.~\ref{fig:schematic} and \ref{fig:enhance}) simply correspond to
irregularities of the emission along the cone.

\subsubsection{Fin-shaped features}
\label{subsub:fin}

{The position of the North fin, above and
beyond the asymptotic direction of the synchrone in Fig.~\ref{fig:fp},
suggest an out-of-plane feature, in turn indicating an out-of-plane
ejection velocity vector. It can be explained as a consequence of the
ejection cone described in the previous section. The cone itself
defines the left edge of the fin, the smaller dust 
being driven back to create the body of the feature.
The SE fin could not be reproduced.
}

\subsection{Alternative emission processes}
\label{subsub:alt}

When evaluating the characteristics of the nucleus, in Section
\ref{sec:thermal}, we concluded that no water ice or more volatile
species would have survived for more than a few million years in the
body on its current orbit. This makes cometary activity a very
unlikely cause for the observed dust emission, as it would require a
mechanism to inject the object on its current Flora-like orbit in the
recent past.

The FP modelling of the tail (Section~\ref{sec:FP}) and other studies
\citep{Jew+10,Sno+10} firmly indicate the emission duration was very
short, in the form of a burst that took place around Feb. 2009.  The
more detailed dynamical dust models presented in the previous section
suggest this emission took place on a cone with an half-opening
$\alpha=40\degr$. As we will discuss in the following section, this
strongly suggests that the dust emission corresponds to an ejecta cone
resulting from an impact with a smaller body. In this section, we
shall briefly consider other alternative dust emission processes:
rotational spin-up and electrostatic dust levitation.

\paragraph{Rotational spin-up}: Small objects in the main belt can be
driven toward rotational instability by radiation
torque \citet{Rub00}. Radiation torque can spin-up or spin-down the
object depending on its shape and the nature of the object, and the
numerous small asteroids that are either fast and slow rotators
suggest this effect is indeed acting widely \citep{PH00}. However, 
simulations indicate that the ejection of material from a spun-up
object occurs along the equatorial disk \citep{Wal+08}, i.e. with a
geometry that is very different from what is observed here. These
numerical simulations were done for particles larger than
dust. However, the geometry of the large grains observed in this
object seems completely incompatible with the rotational spin-up
lifting.

\paragraph{Electrostatic lift:}
Spacecraft have observed fine particles levitated at velocities near
1~m~s$^{-1}$ at the Moon's terminator caused by the buildup of a
potential difference between illuminated and shadowed regions
\citep{deB+77_photoelectric}. On small asteroids, this effect can
launch dust above the local escape velocity \citep{Lee96_levitation},
creating visible mass loss.  However, there are thousands of asteroids
similar in size to the MBCs that do not eject observable
dust. Furthermore, the FP analysis of the dust around P/2010~A2 indicates
it is constituted of large dust grains, which could not be lifted up
by electrostatic force.  


\subsection{Physical implications of ejecta geometry}

One remaining source of the observed release of dust is the excavation
of material from an impact on the nucleus.  In this section, we
demonstrate to first order that an impact event is consistent with,
and could explain, the three main requirements from observations: the
duration of release, velocity distribution and provenance, and amount
of mass.  We restrict our analysis to a canonical vertical impact
here, although impacts at normal incidence rarely occur on planetary objects; instead, the vast majority occur obliquely for both spherical \citep{1893QB591.G46......,1963mmc..book..301S} and asteroidal-type bodies \citep{1999Icar..140...34C}.

In a hypervelocity impact event, strong shockwaves and subsequent
rarefaction waves set material in motion, excavating and ejecting a
portion of the displaced target up and out of the crater;  the
remainder of the displaced mass remains compressed in the target.  In general, the velocities of main-stage
ejecta can be described by a power-law relationship with either time
or launch position, as dimensionally predicted
\citep[e.g.,][]{Housen_1983}, and demonstrated by a number of
experimental
\citep[e.g.,][]{1973E&PSL..20..226M,1979JGR....84.7669S,1980LPI....11..877P,1976LPSC....7.2983O,1977SPIE...97..177P,1999M&PS...34..605C,2003JGRE..108.5094A}
and numerical studies \citep[e.g.][]{Wada2006528}.  The ejection
angles of ejecta are a function of the material properties of the
target.  In granular target materials (like regolith, as expected on
an asteroidal surface), ejection angles (referenced from local
horizontal) $\psi=90\degr-\alpha_{\rm c}$ range from $\sim30\degr$ to
$\sim50\degr$
\citep{1999M&PS...34..605C,Anderson:un,2005LPI....36.1773A,Hermalyn2010866}
with an average $\alpha_{\rm c}\sim 45\degr$ \citep{Housen_1983} and
evolve throughout crater growth.  In cohesive targets (such as solid
basalt or a welded material), ejection angles tend to be considerably
higher \citep[$\psi \sim 70\degr$][]{1973Moon....6...32G}.
Additionally, highly porous material exhibits a high-ejection angle
component \citep{Schultz200784}.  Our finding of an $\alpha_{\rm c} =
40\degr$ hollow cone is in agreement with an impact into an
unconsolidated, granular target.  The progression from lower ejection
angles (larger $\alpha_{c}$) for faster velocity material to higher
ejection angles (smaller $\alpha_{c}$) at later times and slower
velocities is consistent with the ejection angle evolution in
experiments and computations for unconsolidated granular material
\citep{Hermalyn2010866,DensityEffectsINPress2011}.

Crater growth is arrested either when the ejecta velocities are
insufficient to launch material above the crater rim (gravity-controlled)
or when the shock strength drops below the yield strength of the
material (strength-controlled).  The cutoff in velocity solution space
of $\sim 0.10$~m~s$^{-1}$ is coincident with the escape velocity of the
body.  Therefore, any material ejected under this velocity component
could have returned to the surface of P/2010~A2 and no record of this
material would be apparent in the images.  This behavior would be
consistent with a gravity-scaled crater.  Alternatively, if the surface
exhibits a degree of material strength, crater growth could be arrested
with virtually no ejecta launched below the 0.10~m~s$^{-1}$ cutoff.  We
note that although it is impossible to separate the effects of gravity
and strength controlled crater arrestment without watching the evolution of the ejecta curtain.  Since both gravity and
strength are inferred to be weak, both probably play a role in the
termination of growth.  It is possible to define a theoretical gravity-controlled crater radius as a limiting size since ejection velocities
during main-stage growth will be unaffected by the difference in termination processes.  

\begin{figure}
    \includegraphics[width=8.8cm]{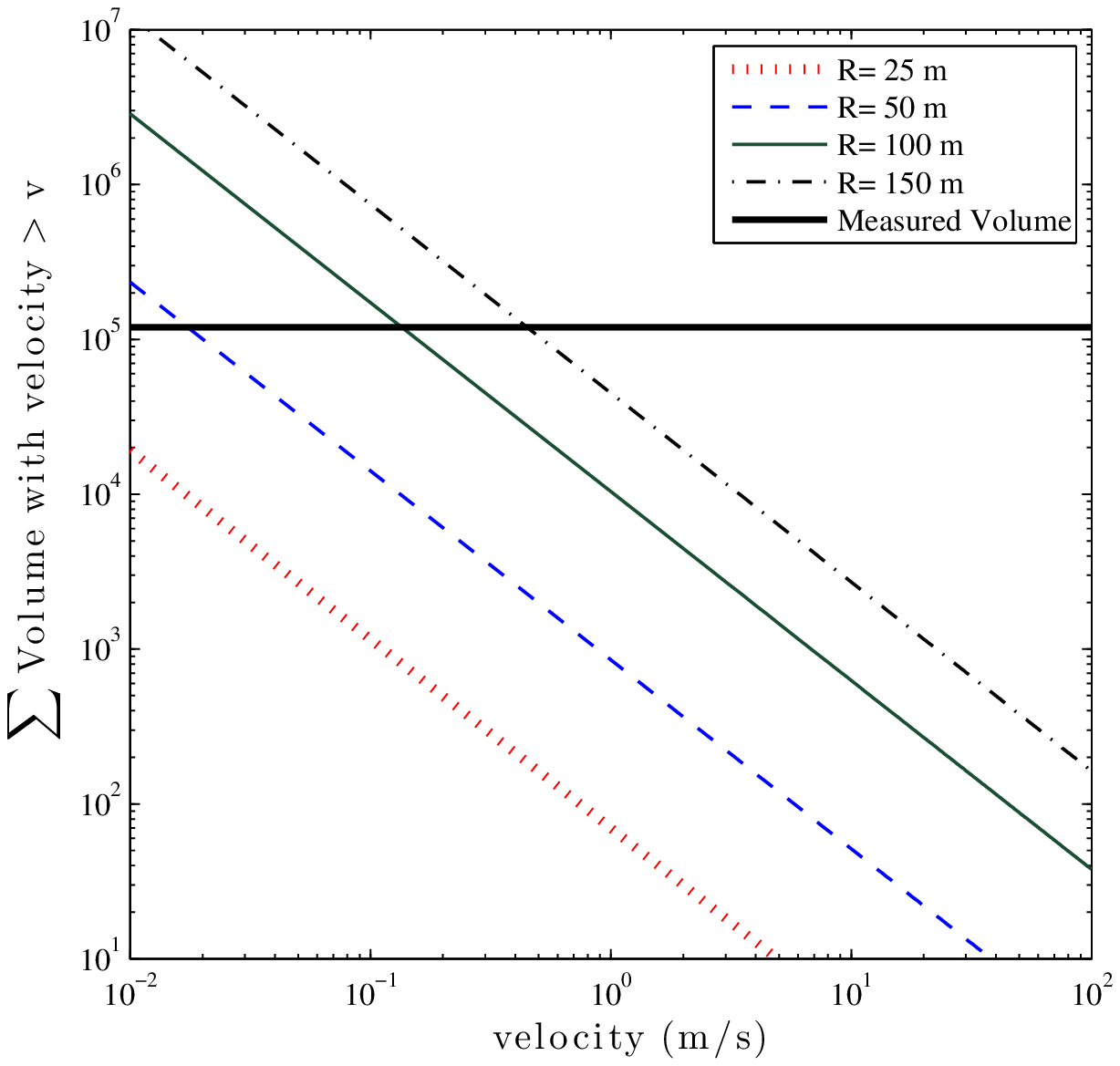}
    \caption{Cumulative volume of ejecta with higher velocity for a set of crater sizes in the gravity-scaled regime.  Measured volume of material is indicated by solid black line.  A crater approximately 100m in radius is required to excavate sufficient material above the escape velocity of the body to match observations.}
    \label{fig:ejecta_volume}
 \end{figure}

In Section~\ref{sec:dust}, we concluded that the visible ejecta corresponds to a sphere with a radius $r_e \sim 40$~m. \citet{Jew+10}
estimated $r_e$ in the 17--36~m range. In both studies, the dust particles were found in the 2--20~mm range, which further supports the excavation of fine regolith-like material in a gravity-scaling regime.  To approach the size of the crater, we look to the ejected mass as a constraint.  The total amount of mass ejected from a crater tends to only be
approximately 50\% of the total displaced mass 
\citep[e.g.][]{1975JGR....80.4062S, 1980LPSC...11.2347C}.  The cumulative (and incremental) ejected mass above a certain velocity similarly follows a power-law relationship with velocity such that the vast majority of material is ejected at low velocities near the crater rim.  The form of this equation for purely gravity-scaled craters is \citep[as given in][]{Housen_1983}:

\begin{equation}
\frac{\sum V_{e}}{R^{3}}=0.32 \left(\frac{v}{\sqrt{g R}}\right)^{-1.22\pm0.02}
\end{equation}

where $V_{e}$ is the cumulative volume of ejecta with higher velocity, $R$ is the final gravity-scaled crater radius, $v$ is the velocity of ejecta, and $g$ is the gravity on the surface; the exponent and coefficient are fit to data from small-scale explosion and impact experiments into granular targets.  Using the higher density values defined above, a crater approximately 100m in radius is required to eject sufficient material above escape velocity to match the observed minimum ejected mass (see Fig. \ref{fig:ejecta_volume}).  The total time to eject the observed material would be on the order of seconds.  A strength-controlled crater follows a  similar relationship (see \citet{Housen_1983} for details), but would require a much larger crater to eject the measured volume of material for anything other than extremely small values of material strength $Y$.  Considerably larger craters are probably not physically possible on such a small body since even the gravity-scaled limiting radius is on the order of the radius of the body itself.  However, there are other bodies that exhibit craters or large-scale depression features of radii approaching that of the body itself; e.g., Mathilde, Phobos and Deimos, Gaspra, and even the Moon.   

One possible mechanism that could help explain the development of such large craters without catastrophically disrupting the body is an oblique impact, which decreases the peak shock pressures while still excavating a large amount of material \citep{1997DPS....29.0309S,1999Icar..140...34C,Schultz-P.H.2011Origin-of-nears}.    This process could also explain the enhancement of higher velocity material ($>0.25$~m~s$^{-1}$) in the images, since ejecta downrange (i.e., in the direction of impact) is enhanced in mass and velocity throughout much of crater growth \citep{2003JGRE..108.5094A} and may appear strongly foreshortened in the images due to perspective. The body may also be fairly porous, which will serve to further attenuate the peak shock pressures.

\section{Summary and discussion}\label{sec:con}

The dust tail of P/2010~A2 was observed and analyzed so to cast some
light on the possible cometary nature of that object, as its
appearance and orbit  could make  
it a member of the small family of Main Belt Comets. The main results
and conclusions of this study are listed
hereafter. 

\begin{itemize}
\item Observations were acquired with GN, NTT, UH~2.2-m from 14
  Jan. until 19 Feb. 2010. 
  The tail did not
  show any measurable evolution over that time span, which included
  the crossing of the orbital plane, which indicates the tail is an
  out-of-plane 3D structure, suggesting the dust was released with a
  significant orthogonal velocity.
\item The nucleus, detached from the tail, appears not to be
  surrounded by dust (at most 3\% of the light corresponds to
  near-nucleus dust); dynamical dust models also indicate that no
  recent dust is present, suggesting no cometary activity at the time
  of the observations. The absolute magnitude of the comet was
  estimated to SDSS $r'(1,1,0)=21.74\pm0.04$ on 19~Jan. 2010 and
  $21.55\pm0.05$ on 2~Feb. Assuming an albedo $p=0.11$, a value typical
  for S-type asteroids, this converts into a nucleus of radius
  $r=80$--90~m. Using a density of $\rho=3$\,000~kg~m$^{-3}$ (also
  typical of a S-type asteroid), the escape velocity of the body is
  $v_e=0.10$--0.12~m~s$^{-1}$.
\item A thermal model of the nucleus indicates that water ice (as well
  as any more volatile ice) would not survive more than a few million
  years in the object on its current orbit. As we have no reason to
  suspect a recent insertion on the current orbit, this rules out
  ice sublimation as the source of the observed dust. Rotational spin
  up and electrostatic lifting were also rejected as possible source
  for the dust tail. 
\item A Finston-Probstein-type dynamical dust modelling of the
  morphology, light distribution and detachment of the tail indicates
  that a short burst of dust emission (a day long or less) represents
  best the observed tail, with an ejection velocity in the range
  $v_e=0.20$--0.30~m~s$^{-1}$. The ejection date is in agreement with
  other studies of this object \citep{Jew+10, Sno+10}; we continue
  this study using Snodgrass' date, 10 Feb. 2009. The model, together
  with direct measurements of the tail within the camera's field of
  view, indicate it is constituted of dust grains with radii $a$
  ranging between a few tenths of millimetres to 20~mm, with a number
  of particles proportional to $a^{-3.44\pm0.08}$ 
{(differential size distribution index).}
  The total dust content of the tail is estimated
  to at least $8\times 10^8$~kg, which could be packed in a sphere
  with a radius of 40~m (assuming the same density
  $\rho=3$\,000~kg~m$^{-3}$ as above).
\item The position and shape of the tail's envelope was modeled as
  originating from dust emitted on a cone with a half-opening angle
  $\alpha \sim 40\degr$ pointing below the orbital plane, toward the
  forward direction, with a ejection velocity $v_e \le
  0.20$~m~s$^{-1}$. The X-shaped arcs were modeled as a
  collection of large particles, with $v_e \ge 0.20$~m~s$^{-1}$; they
  were independently found to be emitted on a section of the same
  hollow cone. The are seen as an arc as a consequence of the
  geometric foreshortening of the cone section (after one year of
  evolution); other sections of the cone, which are not foreshortened
  in the same way, contribute to the overall tail. The Rosetta
  observations \citep{Sno+10} did not see other sections of the cone
  as arcs because of the coarser resolution of its camera.
\item Because ice sublimation, electrostatic lifting and rotational
  spin up were rejected as possible causes for the dust emission, and
  because of the very short release time and the geometry of the dust
  emission on a hollow cone with an half-opening angle $\alpha =
  40\degr$, we conclude that the dust release was caused by an
  impact by a small object. 
\item Considering the volume of ejected dust in the framework of
  a gravity-dominated crater formation process, a crater of $\sim
  100$~m radius was produced in a total time of the order of
  seconds. While that crater is large compared to the body, other
  bodies show craters as large as themselves. An oblique impact
  ---which is statistically the most likely--- would explain how
  P/2010~A2 escaped complete, catastrophic disruption.

\end{itemize}

In summary, the tail of P/2010~A2 can be explained by an impact on a
$\sim 80$--90~m asteroid. The impact released dust in a hollow cone,
which evolved for about one year to form the observed dust
distribution.  The fine-grained dust formed a long tail that was
extended by solar radiation pressure, and the large grains
(un-affected by the solar radiation pressure) retained the conical
shape. The viewing geometry revealed sections of that cone as the arcs
in the head of the dust cloud. Therefore, what we witnessed was likely
an event of collisional grinding within the Flora collisional family,
and not a genuine, water-ice driven Main Belt Comet. 
With increasingly sensitive sky surveys covering larger fractions of
the sky during each dark period, we can expect to detect more and more
of these collisions, as illustrated by the recent discovery of the
dust cloud around (596)~Scheila. While the original goal of this
project was to probe the water ice content in the asteroid main belt,
its end result was the observation of a natural impact on a small
asteroid.


\begin{acknowledgements}
We are grateful to director of the Gemini Observatory for the very
rapid allocation of discretionary time to this project, to Hermann
Boehnhardt (MPI), Javier Licandro (IAC) and GianPaolo Tozzi
(Arcetri) for very useful discussions on this object, 
to Heather Kaluna (IfA) for acquiring some of the data presented, and
to Colin Snodgrass who refereed this paper.

Image processing in this paper has been performed, in part, using the
IRAF \citep{Tody86} and ESO-MIDAS.  The IRAF software is distributed
by the National Optical Astronomy Observatories, which is operated by
the Association of Universities for Research in Astronomy, Inc. (AURA)
under cooperative agreement with the National Science
Foundation, USA. ESO-MIDAS (version 09SEPpl1.2) was developed and is
distributed by the European Southern Observatory.

This work is based in part on observations obtained at the Gemini
Observatory, which is operated by the Association of Universities for
Research in Astronomy, Inc., under a cooperative agreement with the
NSF on behalf of the Gemini partnership: the National Science
Foundation (United States), the Science and Technology Facilities
Council (United Kingdom), the National Research Council (Canada),
CONICYT (Chile), the Australian Research Council (Australia),
Ministério da Ciência e Tecnologia (Brazil) and Ministerio de Ciencia,
Tecnología e Innovación Productiva (Argentina).

This material is based upon work supported by the National Aeronautics
and Space Administration through the NASA Astrobiology Institute under
Cooperative Agreement No. NNA09DA77A issued through the Office of
Space Science, and by NASA Grant No. NNX07AO44G.

\end{acknowledgements}

\bibliographystyle{apalike} 
\bibliography{mnemonic,P2010A2}

\begin{thebibliography}{90}
\expandafter\ifx\csname natexlab\endcsname\relax\def\natexlab#1{#1}\fi

\bibitem[{{Agarwal} {et~al.}(2007){Agarwal}, {M{\"u}ller}, \&
  {Gr{\"u}n}}]{Aga+07}
{Agarwal}, J., {M{\"u}ller}, M., \& {Gr{\"u}n}, E. 2007, \ssr, 128, 79

\bibitem[{{A'Hearn} {et~al.}(1984){A'Hearn}, {Schleicher}, {Millis}, {Feldman},
  \& {Thompson}}]{AHe+84}
{A'Hearn}, M.~F., {Schleicher}, D.~G., {Millis}, R.~L., {Feldman}, P.~D., \&
  {Thompson}, D.~T. 1984, \aj, 89, 579

\bibitem[{{Anderson} \& {Schultz}(2005)}]{2005LPI....36.1773A}
{Anderson}, J.~L.~B. \& {Schultz}, P.~H. 2005, in Lunar and Planetary Institute
  Science Conference Abstracts, Vol.~36, 36th Annual Lunar and Planetary
  Science Conference, ed. {S.~Mackwell \& E.~Stansbery}, 1773--+

\bibitem[{{Anderson} {et~al.}(2003){Anderson}, {Schultz}, \&
  {Heineck}}]{2003JGRE..108.5094A}
{Anderson}, J.~L.~B., {Schultz}, P.~H., \& {Heineck}, J.~T. 2003, JGR(Planets),
  108, 5094

\bibitem[{{Anderson} {et~al.}(2004){Anderson}, {Schultz}, \&
  {Heineck}}]{Anderson:un}
{Anderson}, J. L.~B., {Schultz}, P.~H., \& {Heineck}, J.~T. 2004, Meteoritics
  \& Planetary Science, 39, 303

\bibitem[{{Birtwhistle} {et~al.}(2010{\natexlab{a}}){Birtwhistle}, {Ryan},
  {Sato}, {Beshore}, \& {Kadota}}]{CBET2114}
{Birtwhistle}, P., {Ryan}, W.~H., {Sato}, H., {Beshore}, E.~C., \& {Kadota}, K.
  2010{\natexlab{a}}, Central Bureau Electronic Telegrams, 2114, 1

\bibitem[{{Birtwhistle} {et~al.}(2010{\natexlab{b}}){Birtwhistle}, {Ryan},
  {Sato}, {Beshore}, \& {Kadota}}]{IAUC9105}
{Birtwhistle}, P., {Ryan}, W.~H., {Sato}, H., {Beshore}, E.~C., \& {Kadota}, K.
  2010{\natexlab{b}}, \iaucirc, 9105, 1

\bibitem[{{Bodewits} {et~al.}(2011){Bodewits}, {Kelley}, {Li}, {Landsman},
  {Besse}, \& {A'Hearn}}]{Bod+11}
{Bodewits}, D., {Kelley}, M.~S., {Li}, J.-Y., {et~al.} 2011, \apjl, 733, L3

\bibitem[{{Britt} {et~al.}(2002){Britt}, {Yeomans}, {Housen}, \&
  {Consolmagno}}]{Bri+02}
{Britt}, D.~T., {Yeomans}, D., {Housen}, K., \& {Consolmagno}, G. 2002,
  Asteroids III, 485

\bibitem[{{Burns} {et~al.}(1979){Burns}, {Lamy}, \& {Soter}}]{BLS79}
{Burns}, J.~A., {Lamy}, P.~L., \& {Soter}, S. 1979, \icarus, 40, 1

\bibitem[{{Buzzoni} {et~al.}(1984){Buzzoni}, {Delabre}, {Dekker}, {Dodorico},
  {Enard}, {Focardi}, {Gustafsson}, {Nees}, {Paureau}, \&
  {Reiss}}]{1984Msngr..38....9B}
{Buzzoni}, B., {Delabre}, B., {Dekker}, H., {et~al.} 1984, The Messenger, 38, 9

\bibitem[{{Cheng} \& {Barnouin-Jha}(1999)}]{1999Icar..140...34C}
{Cheng}, A.~F. \& {Barnouin-Jha}, O.~S. 1999, \icarus, 140, 34

\bibitem[{{Cintala} {et~al.}(1999){Cintala}, {Berthoud}, \&
  {H{\"o}rz}}]{1999M&PS...34..605C}
{Cintala}, M.~J., {Berthoud}, L., \& {H{\"o}rz}, F. 1999, Meteoritics and
  Planetary Science, 34, 605

\bibitem[{{Cohen} \& {Coker}(2000)}]{Cohen+00}
{Cohen}, B.~A. \& {Coker}, R.~F. 2000, \icarus, 145, 369

\bibitem[{{Croft}(1980)}]{1980LPSC...11.2347C}
{Croft}, S.~K. 1980, in Lunar and Planetary Science Conference Proceedings,
  Vol.~11, Lunar and Planetary Science Conference Proceedings, ed.
  {S.~A.~Bedini}, 2347--2378

\bibitem[{{de Bibhas} \& {Criswell}(1977)}]{deB+77_photoelectric}
{de Bibhas}, R. \& {Criswell}, D. 1977, JGR, 82, 999–1004

\bibitem[{{Delsemme}(1982)}]{Del82}
{Delsemme}, A.~H. 1982, in IAU Colloq. 61: Comet Discoveries, Statistics, and
  Observational Selection, ed. {L.~L.~Wilkening}, 85--130

\bibitem[{{Dohnanyi}(1969)}]{Doh69}
{Dohnanyi}, J.~S. 1969, \jgr, 74, 2531

\bibitem[{{Encrenaz}(2008)}]{Encrenaz08}
{Encrenaz}, T. 2008, \araa, 46, 57

\bibitem[{{Farnham}(1996)}]{farnham96}
{Farnham}, T.~L. 1996, PhD thesis, University if Hawai`i

\bibitem[{{Finson} \& {Probstein}(1968)}]{FP68}
{Finson}, M. \& {Probstein}, R. 1968, \apj, 154, 327

\bibitem[{{Florczak} {et~al.}(1998){Florczak}, {Barucci}, {Doressoundiram},
  {Lazzaro}, {Angeli}, \& {Dotto}}]{Flo+98}
{Florczak}, M., {Barucci}, M.~A., {Doressoundiram}, A., {et~al.} 1998, \icarus,
  133, 233

\bibitem[{{Fukugita} {et~al.}(1996){Fukugita}, {Ichikawa}, {Gunn}, {Doi},
  {Shimasaku}, \& {Schneider}}]{1996AJ....111.1748F}
{Fukugita}, M., {Ichikawa}, T., {Gunn}, J.~E., {et~al.} 1996, \aj, 111, 1748

\bibitem[{{Garaud} \& {Lin}(2007)}]{GaraudL07_snow}
{Garaud}, P. \& {Lin}, D.~N.~C. 2007, \apj, 654, 606

\bibitem[{{Gault}(1973)}]{1973Moon....6...32G}
{Gault}, D.~E. 1973, Moon, 6, 32

\bibitem[{{Genda} \& {Ikoma}(2007)}]{GI07_DHratio}
{Genda}, H. \& {Ikoma}, M. 2007, in Planetary Atmospheres, 43

\bibitem[{{Genda} \& {Ikoma}(2008)}]{GI08_ocean}
{Genda}, H. \& {Ikoma}, M. 2008, \icarus, 194, 42

\bibitem[{{Gilbert}(1893)}]{1893QB591.G46......}
{Gilbert}, G.~K. 1893, {The moon's face. A study of the origin of its
  features.}, Vol. Bulletin XII (Philosophical Society of Washington)

\bibitem[{{Green}(2010)}]{IAUC9109}
{Green}, D.~W.~E. 2010, \iaucirc, 9109, 1

\bibitem[{{Grimm} \& {McSween}(1989)}]{Grimm+89}
{Grimm}, R.~E. \& {McSween}, Jr., H.~Y. 1989, \icarus, 82, 244

\bibitem[{{Haghighipour}(2009)}]{Haghighipour09_MBC}
{Haghighipour}, N. 2009, in ``Icy Bodies of the Solar System,'' Proc. IAU Symp.
  263, J. Fernandez et al., Ed.

\bibitem[{{Haver} {et~al.}(2010){Haver}, {Caradossi}, \& {Buzzi}}]{CBET2134}
{Haver}, R., {Caradossi}, A., \& {Buzzi}, L. 2010, Central Bureau Electronic
  Telegrams, 2134, 4

\bibitem[{Hermalyn \& Schultz(2010)}]{Hermalyn2010866}
Hermalyn, B. \& Schultz, P.~H. 2010, Icarus, 209, 866

\bibitem[{{Hermalyn} \& {Schultz}(2011)}]{DensityEffectsINPress2011}
{Hermalyn}, B. \& {Schultz}, P.~H. 2011, Icarus, In press, dOI:
  10.1016/j.icarus.2011.09.008

\bibitem[{Hermalyn {et~al.}(2011)Hermalyn, Schulz, {Hainaut}, \&
  {Meech}}]{PAPER3}
Hermalyn, B., Schulz, P., {Hainaut}, O.~R., \& {Meech}, K.~J. 2011, Astron.
  Astrophys., in preparation

\bibitem[{{Hook} {et~al.}(2004){Hook}, {J{\o}rgensen}, {Allington-Smith},
  {Davies}, {Metcalfe}, {Murowinski}, \& {Crampton}}]{2004PASP..116..425H}
{Hook}, I.~M., {J{\o}rgensen}, I., {Allington-Smith}, J.~R., {et~al.} 2004,
  \pasp, 116, 425

\bibitem[{{Housen} {et~al.}(1983){Housen}, {Schmidt}, \&
  {Holsapple}}]{Housen_1983}
{Housen}, K.~R., {Schmidt}, R.~M., \& {Holsapple}, K.~A. 1983, JGR, 88, 2485

\bibitem[{{Hsieh}(2009)}]{HHH09}
{Hsieh}, H.~H. 2009, \aap, 505, 1297

\bibitem[{{Hsieh} \& {Jewitt}(2006)}]{HsiehJ06_MBC}
{Hsieh}, H.~H. \& {Jewitt}, D. 2006, Science, 312, 561

\bibitem[{{Ivezi{\'c}} {et~al.}(2001){Ivezi{\'c}}, {Tabachnik}, {Rafikov},
  {Lupton}, {Quinn}, {Hammergren}, {Eyer}, {Chu}, {Armstrong}, {Fan},
  {Finlator}, {Geballe}, {Gunn}, {Hennessy}, {Knapp}, {Leggett}, {Munn},
  {Pier}, {Rockosi}, {Schneider}, {Strauss}, {Yanny}, {Brinkmann}, {Csabai},
  {Hindsley}, {Kent}, {Lamb}, {Margon}, {McKay}, {Smith}, {Waddel}, {York}, \&
  {the SDSS Collaboration}}]{2001AJ....122.2749I}
{Ivezi{\'c}}, {\v Z}., {Tabachnik}, S., {Rafikov}, R., {et~al.} 2001, \aj, 122,
  2749

\bibitem[{{Jedicke} {et~al.}(2002){Jedicke}, {Larsen}, \& {Spahr}}]{JLS02}
{Jedicke}, R., {Larsen}, J., \& {Spahr}, T. 2002, Asteroids III, 71

\bibitem[{{Jewitt} {et~al.}(2010){Jewitt}, {Weaver}, {Agarwal}, {Mutchler}, \&
  {Drahus}}]{Jew+10}
{Jewitt}, D., {Weaver}, H., {Agarwal}, J., {Mutchler}, M., \& {Drahus}, M.
  2010, \nat, 467, 817

\bibitem[{{Jewitt} {et~al.}(2011){Jewitt}, {Weaver}, {Mutchler}, {Larson}, \&
  {Agarwal}}]{Jew+11}
{Jewitt}, D., {Weaver}, H., {Mutchler}, M., {Larson}, S., \& {Agarwal}, J.
  2011, \apjl, 733, L4

\bibitem[{{Jones} {et~al.}(1990){Jones}, {Lebofsky}, {Lewis}, \&
  {Marley}}]{Jon+90}
{Jones}, T.~D., {Lebofsky}, L.~A., {Lewis}, J.~S., \& {Marley}, M.~S. 1990,
  \icarus, 88, 172

\bibitem[{{Kasting} \& {Catling}(2003)}]{KC03_habitable}
{Kasting}, J.~F. \& {Catling}, D. 2003, \araa, 41, 429

\bibitem[{{Kearsley} {et~al.}(2009){Kearsley}, {Burchell}, {Price}, {Graham},
  {Wozniakiewicz}, {Cole}, {Foster}, \& {Teslich}}]{2009M&PS...44.1489K}
{Kearsley}, A.~T., {Burchell}, M.~J., {Price}, M.~C., {et~al.} 2009,
  Meteoritics and Planetary Science, 44, 1489

\bibitem[{{Kleyna} {et~al.}(2011){Kleyna}, {Hainaut}, {Zenn}, \&
  {Meech}}]{PAPER2}
{Kleyna}, J., {Hainaut}, O.~R., {Zenn}, A., \& {Meech}, K.~J. 2011, Astron.
  Astrophys., submitted

\bibitem[{{Kresak}(1980)}]{Kresak80}
{Kresak}, L. 1980, Moon and Planets, 22, 83

\bibitem[{{Krot} {et~al.}(2006){Krot}, {Hutcheon}, {Brearley}, {Pravdivtseva},
  {Petaev}, \& {Hohenberg}}]{Krot+06}
{Krot}, A.~N., {Hutcheon}, I.~D., {Brearley}, A.~J., {et~al.} 2006, {Timescales
  and Settings for Alteration of Chondritic Meteorites} ({Lauretta, D.~S.~\&
  McSween, H.~Y.}), 525--553

\bibitem[{Landolt(1992)}]{landolt92}
Landolt, A. 1992, Astrophys. J., 104, 340

\bibitem[{{Larson}(2010)}]{Lar10}
{Larson}, S.~M. 2010, \iaucirc, 9188, 1

\bibitem[{{Lecar} {et~al.}(2006){Lecar}, {Podolak}, {Sasselov}, \&
  {Chiang}}]{Lec+06}
{Lecar}, M., {Podolak}, M., {Sasselov}, D., \& {Chiang}, E. 2006, \apj, 640,
  1115

\bibitem[{{Lee}(1996)}]{Lee96_levitation}
{Lee}, P. 1996, \icarus, 124, 181

\bibitem[{{Machida} \& {Abe}(2010)}]{Mac+10}
{Machida}, R. \& {Abe}, Y. 2010, \apj, 716, 1252

\bibitem[{{Marsden} {et~al.}(2010){Marsden}, {Birtwhistle}, {Holmes}, {Foglia},
  {Scotti}, {Sostero}, {Guido}, {Donato}, \& {Gonano}}]{Mar+10}
{Marsden}, B.~G., {Birtwhistle}, P., {Holmes}, R., {et~al.} 2010, \iaucirc,
  9169, 1

\bibitem[{{McGetchin} {et~al.}(1973){McGetchin}, {Settle}, \&
  {Head}}]{1973E&PSL..20..226M}
{McGetchin}, T.~R., {Settle}, M., \& {Head}, J.~W. 1973, Earth and Planetary
  Science Letters, 20, 226

\bibitem[{{Meech} \& {Svoren}(2004)}]{2004come.book..317M}
{Meech}, K.~J. \& {Svoren}, J. 2004, {Using cometary activity to trace the
  physical and chemical evolution of cometary nuclei} ({Festou, M.~C., Keller,
  H.~U., \& Weaver, H.~A.}), 317--335

\bibitem[{{Michikami} {et~al.}(2008){Michikami}, {Nakamura}, {Hirata},
  {Gaskell}, {Nakamura}, {Honda}, {Honda}, {Hiraoka}, {Saito}, {Demura},
  {Ishiguro}, \& {Miyamoto}}]{Nak+08}
{Michikami}, T., {Nakamura}, A.~M., {Hirata}, N., {et~al.} 2008, Earth,
  Planets, and Space, 60, 13

\bibitem[{{Min} {et~al.}(2011){Min}, {Dullemond}, {Kama}, \&
  {Dominik}}]{Min+11}
{Min}, M., {Dullemond}, C.~P., {Kama}, M., \& {Dominik}, C. 2011, \icarus, 212,
  416

\bibitem[{{Morbidelli} {et~al.}(2000){Morbidelli}, {Chambers}, {Lunine},
  {Petit}, {Robert}, {Valsecchi}, \& {Cyr}}]{Mor+00_water}
{Morbidelli}, A., {Chambers}, J., {Lunine}, J.~I., {et~al.} 2000, Meteoritics
  and Planetary Science, 35, 1309

\bibitem[{{Moreno}(2009)}]{M09}
{Moreno}, F. 2009, \apjs, 183, 33

\bibitem[{{Moreno} {et~al.}(2010){Moreno}, {Licandro}, {Tozzi}, {Ortiz},
  {Cabrera-Lavers}, {Augusteijn}, {Liimets}, {Lindberg}, {Pursimo},
  {Rodr{\'{\i}}guez-Gil}, \& {Vaduvescu}}]{Mor+10}
{Moreno}, F., {Licandro}, J., {Tozzi}, G., {et~al.} 2010, \apjl, 718, L132

\bibitem[{{Mottl} {et~al.}(2007){Mottl}, {Glazer}, {Kaiser}, \&
  {Meech}}]{Mot+07_water}
{Mottl}, M., {Glazer}, B., {Kaiser}, R., \& {Meech}, K. 2007, Chemie der Erde /
  Geochemistry, 67, 253

\bibitem[{{Muralidharan} {et~al.}(2008){Muralidharan}, {Deymier}, {Stimpfl},
  {de Leeuw}, \& {Drake}}]{Mur+08_water}
{Muralidharan}, K., {Deymier}, P., {Stimpfl}, M., {de Leeuw}, N.~H., \&
  {Drake}, M.~J. 2008, \icarus, 198, 400

\bibitem[{{Oberbeck} \& {Morrison}(1976)}]{1976LPSC....7.2983O}
{Oberbeck}, V.~R. \& {Morrison}, R.~H. 1976, in Lunar and Planetary Science
  Conference Proceedings, ed. R.~B. {Merrill}, Vol.~7, 2983--3005

\bibitem[{{Piekutowski}(1980)}]{1980LPI....11..877P}
{Piekutowski}, A.~J. 1980, in Lunar and Planetary Inst. Technical Report,
  Vol.~11, Lunar and Planetary Institute Science Conference Abstracts, 877--878

\bibitem[{{Piekutowski} {et~al.}(1977){Piekutowski}, {Andrews}, \&
  {Swift}}]{1977SPIE...97..177P}
{Piekutowski}, A.~J., {Andrews}, R.~J., \& {Swift}, H.~F. 1977, in Society of
  Photo-Optical Instrumentation Engineers (SPIE) Conference Series, ed. M.~C.
  {Richardson}, Vol.~97

\bibitem[{{Pravec} \& {Harris}(2000)}]{PH00}
{Pravec}, P. \& {Harris}, A.~W. 2000, \icarus, 148, 12

\bibitem[{{Prialnik}(1992)}]{1992ApJ...388..196P}
{Prialnik}, D. 1992, \apj, 388, 196

\bibitem[{{Prialnik} {et~al.}(2004){Prialnik}, {Benkhoff}, \&
  {Podolak}}]{2004come.book..359P}
{Prialnik}, D., {Benkhoff}, J., \& {Podolak}, M. 2004, {Modeling the structure
  and activity of comet nuclei} ({Festou, M.~C., Keller, H.~U., \& Weaver,
  H.~A.}), 359--387

\bibitem[{{Prialnik} \& {Rosenberg}(2009)}]{2009MNRAS.399L..79P}
{Prialnik}, D. \& {Rosenberg}, E.~D. 2009, \mnras, 399, L79

\bibitem[{{Prialnik} {et~al.}(2008){Prialnik}, {Sarid}, {Rosenberg}, \&
  {Merk}}]{2008SSRv..138..147P}
{Prialnik}, D., {Sarid}, G., {Rosenberg}, E.~D., \& {Merk}, R. 2008, \ssr, 138,
  147

\bibitem[{{Rivkin} {et~al.}(2002){Rivkin}, {Howell}, {Vilas}, \&
  {Lebofsky}}]{Riv+02}
{Rivkin}, A.~S., {Howell}, E.~S., {Vilas}, F., \& {Lebofsky}, L.~A. 2002,
  Asteroids III, 235

\bibitem[{{Rubincam}(2000)}]{Rub00}
{Rubincam}, D.~P. 2000, \icarus, 148, 2

\bibitem[{{Sarid} {et~al.}(2005){Sarid}, {Prialnik}, {Meech}, {Pittichov{\'a}},
  \& {Farnham}}]{2005PASP..117..796S}
{Sarid}, G., {Prialnik}, D., {Meech}, K.~J., {Pittichov{\'a}}, J., \&
  {Farnham}, T.~L. 2005, \pasp, 117, 796

\bibitem[{{Schultz}(1997)}]{1997DPS....29.0309S}
{Schultz}, P.~H. 1997, in Bulletin of the American Astronomical Society,
  Vol.~29, AAS/Division for Planetary Sciences Meeting Abstracts \#29, 960--+

\bibitem[{Schultz {et~al.}(2007)Schultz, Eberhardy, Ernst, A'Hearn, Sunshine,
  \& Lisse}]{Schultz200784}
Schultz, P.~H., Eberhardy, C.~A., Ernst, C.~M., {et~al.} 2007, Icarus, 191, 84
  , deep Impact at Comet Tempel 1

\bibitem[{{Schultz} \& {Gault}(1979)}]{1979JGR....84.7669S}
{Schultz}, P.~H. \& {Gault}, D.~E. 1979, JGR, 84, 7669

\bibitem[{{Schultz, P.~H.} \&
  {Crawford}(2011)}]{Schultz-P.H.2011Origin-of-nears}
{Schultz, P.~H.} \& {Crawford}, D.~A. 2011, Geological Society of America
  Special Papers, 477, 141

\bibitem[{{Shoemaker}(1963)}]{1963mmc..book..301S}
{Shoemaker}, E.~M. 1963, {Impact Mechanics at Meteor Crater, Arizona} ({Kuiper,
  G.~P.~\& Middlehurst, B.~M.}), 301--+

\bibitem[{{Snodgrass} {et~al.}(2008){Snodgrass}, {Saviane}, {Monaco}, \&
  {Sinclaire}}]{2008Msngr.132...18S}
{Snodgrass}, C., {Saviane}, I., {Monaco}, L., \& {Sinclaire}, P. 2008, The
  Messenger, 132, 18

\bibitem[{{Snodgrass} {et~al.}(2010){Snodgrass}, {Tubiana}, {Vincent},
  {Sierks}, {Hviid}, {Moissi}, {Boehnhardt}, {Barbieri}, {Koschny}, {Lamy},
  {Rickman}, {Rodrigo}, {Carry}, {Lowry}, {Laird}, {Weissman}, {Fitzsimmons},
  {Marchi}, \& {OSIRIS Team}}]{Sno+10}
{Snodgrass}, C., {Tubiana}, C., {Vincent}, J., {et~al.} 2010, \nat, 467, 814

\bibitem[{{St{\"o}effler} {et~al.}(1975){St{\"o}effler}, {Gault}, {Wedekind},
  \& {Polkowski}}]{1975JGR....80.4062S}
{St{\"o}effler}, D., {Gault}, D.~E., {Wedekind}, J., \& {Polkowski}, G. 1975,
  JGR, 80, 4062

\bibitem[{{Tody}(1986)}]{Tody86}
{Tody}, D. 1986, in ``SPIE Instrumentation in Astronomy VI,'' Ed. D. L.
  Crawford, Vol. 627, 733--748

\bibitem[{{Vaghi}(1973)}]{Vaghi73}
{Vaghi}, S. 1973, \aap, 29, 85

\bibitem[{Wada {et~al.}(2006)Wada, Senshu, \& Matsui}]{Wada2006528}
Wada, K., Senshu, H., \& Matsui, T. 2006, Icarus, 180, 528

\bibitem[{{Wainscoat}(2010)}]{UHCCD}
{Wainscoat}, R.~J. 2010, University of Hawaii 2.2-meter telescope Observer
  information, World Wide Web electronic publication, retreived on 2010-Jun-10

\bibitem[{{Walsh} {et~al.}(2008){Walsh}, {Richardson}, \& {Michel}}]{Wal+08}
{Walsh}, K.~J., {Richardson}, D.~C., \& {Michel}, P. 2008, \nat, 454, 188

\bibitem[{{Weidenschilling}(2004)}]{2004come.book...97W}
{Weidenschilling}, S.~J. 2004, {From icy grains to comets} ({Festou, M.~C.,
  Keller, H.~U., \& Weaver, H.~A.}), 97--104

\bibitem[{{Wilson} {et~al.}(1999){Wilson}, {Keil}, {Browning}, {Krot}, \&
  {Bourcier}}]{Wilson+99}
{Wilson}, L., {Keil}, K., {Browning}, L.~B., {Krot}, A.~N., \& {Bourcier}, W.
  1999, Meteoritics and Planetary Science, 34

\end{thebibliography}




\end{document}